# Robust multicellular programs dissect the complex tumor microenvironment and track disease progression in colorectal adenocarcinomas

Loan Vulliard[1,2], Teresa Glauner[1], Sven Truxa[1,3], Miray Cetin[1,3], Yu-Le Wu[1], Ronald Simon[4], Laura Behm[4], Jovan Tanevski[2,5], Julio Saez-Rodriguez[2,5,6], Guido Sauter[4], Felix J. Hartmann[1,7]

[1] Systems Immunology and Single-Cell Biology, German Cancer Research Center (DKFZ), Heidelberg, Germany
[2] Institute for Computational Biomedicine, Heidelberg University, Faculty of Medicine, and Heidelberg University Hospital, Heidelberg, Germany
[3] Heidelberg University, Faculty of Biosciences, Heidelberg, Germany
[4] Institute of Pathology, University Medical Center Hamburg-Eppendorf, Hamburg, Germany
[5] Translational Spatial Profiling Center, Heidelberg University Hospital, Heidelberg, Germany
[6] European Molecular Biology Laboratory, European Bioinformatics Institute (EMBL-EBI), Hinxton, UK
[7] German Cancer Consortium (DKTK), Heidelberg, Germany
* Corresponding authors: felix.hartmann@dkfz-heidelberg.de

## Abstract

Colorectal cancer (CRC) is highly heterogeneous, with five-year survival rates dropping from ~90% in localized disease to ~15% with distant metastases. Disease progression is shaped not only by tumor-intrinsic alterations but also by the reorganization of the tumor microenvironment (TME). Metabolic, compositional, and spatial changes contribute to this progression, but considered individually they lack context and often fail as therapeutic targets. Understanding their coordination could reveal processes to alter the disease course. Here, we combined multiplexed ion beam imaging (MIBI) with machine learning to profile metabolic, functional and spatial states of 522 colorectal lesions with single-cell resolution.

We observed recurrent stage-specific remodeling marked by a lymphoid-to-myeloid shift, stromal–cancer cooperation, and malignant metabolic shifts. Spatial organization of epithelial, stromal, and immune compartments provided stronger stratification of disease stage than tumor-intrinsic changes or bulk immune infiltration alone. To systematically model these coordinated changes, we condensed multimodal features into 10 latent factors of TME organization. These factors tracked disease progression, were conserved across cohorts, and revealed frequent multicellular metabolic niches and distinct, non-exclusive TME trajectories.

Our framework MuVIcell exposes the elements that together drive CRC progression by grouping co-occurring changes across cell types and feature classes into coordinated multicellular programs. This creates a rational basis to therapeutically target TME reorganization. Importantly, the framework is scalable and flexible, offering a resource for studying multicellular organization in other solid tumors.

## 1. Introduction

Despite available therapeutic options, CRC remains difficult to treat because of its complexity and heterogeneity[1,2], especially for late diagnoses[3]. A central driver of this challenge is the tumor microenvironment (TME), a dynamic ecosystem of malignant, immune, and stromal cells[4,5], known to influence tumor growth and treatment response[6,7].

Distinct TME neighborhoods and functional units have been characterized in CRC based on patterns of cell type composition, gene expression programs and local interactions[6,8,9]. They comprise regions enriched for Wnt signaling, angiogenesis, or epithelial-mesenchymal transition pathways, and various immune infiltration patterns. Both the phenotypic state and position of immune cells within the TME can influence tumor behavior. For instance, M2-like macrophages can act as pro-tumor signals[10], while CD8+ T cells are found to be more active near tumor cells[4,11]. Cancer-associated fibroblasts (CAF) interact with tumor cells both directly, by supporting their growth structurally and functionally, and indirectly, for example, by secreting inflammatory signals to recruit myeloid cells[9,12]. In parallel, metabolic rewiring, including in glycolysis, oxidative phosphorylation and lipid synthesis, emerges as a hallmark of CRC, reflecting adaptations to hypoxia and nutrient deprivation and reshaping stromal and immune compartments[7].

CRC profiling studies have so far been limited in scope and resolution, preventing a comprehensive view of intratumor heterogeneity and cell–cell interactions. Molecular features are often reported in isolation, without complete mechanistic context or direct connection to oncogenic trajectories[8]. This contributes to the limited clinical translation of candidate targets and biomarkers[13]. In practice, drug mechanisms of action are often multifaceted[14,15], and their effects may be confounded by cellular interactions and disease heterogeneity[16,17]. As a result, individual parameters rarely suffice to guide treatment. For example, response to immune checkpoint inhibitors is shaped by tumor aneuploidy[18], mismatch repair status[19], immunophenotype[20], and mutational burden[21], each of which is indicative but not fully predictive in isolation. More broadly, the coordination of multicellular and metabolic programs within spatial niches and their evolution during CRC progression is key to improve therapeutic targeting. Thus, we aim to systematically map spatial and metabolic heterogeneity in CRC, identify coordinated multicellular programs, and link them to disease progression. By considering tissues as multicellular communities, we can move beyond atlases of isolated cells and capture synchronized TME changes explaining the poor outcomes observed in advanced tumors[27].

The pTNM stages capture key steps in tumor development as evaluated by pathologists and strongly predict patient survival, and can serve as anchors to identify TME features as hallmarks of CRC progression[3,25,26]. A parallel effort to structure CRC heterogeneity was the establishment of the consensus molecular subtypes (CMS) classification, which groups tumors based on recurrent gene expression changes with prognostic relevance[22]. While valuable for precision medicine, CMS subtypes are not directly linked to progression[23]. CMS classes were also found to form spatially distinct communities within individual tumors, highlighting their incompleteness in describing whole CRC lesions through bulk measurements[24]. No comparable attempt has been made to comprehensively characterize recurrent patterns beyond transcriptomic data across the TME.

Here, we address these challenges with comprehensive profiles of the CRC landscape, using multiplexed ion beam imaging by time of flight (MIBI-TOF)[28] to study a cohort of 522 patients. We perform a systematic analysis of the organization of molecules, cells and tissues underlying the disease course using a supervised machine learning approach. Then, we contextualize all the aspects identified as relevant to the TME trajectory, by looking at their coordination using factor analysis and cross-predictions. Finally, we validate the resulting multicellular programs using external CRC cohorts profiled with single-cell sequencing and spatial transcriptomics.

Our approach revealed changes in composition, metabolic and functional states, and spatial organization between pT stages. Major shifts included lymphoid to myeloid rebalancing, malignant metabolic signatures with higher activity across pathways, and stromal support of advanced tumors. These changes were not happening independently but regulated via complex multicellular programs, summarized as 10 latent factors, two of which were especially mapped to pT stages. They corresponded to a tumor-CAF cooperation pattern with reduced FAO, and myeloid-rich glycolytic milieu with remodeled vasculature.

Our quantitative and systematic analysis pointed out the most common axes of variation in CRC progression, which may serve as robust biomarkers or stratification axes for improved CRC care. Furthermore, our framework to extract multicellular programs from single-cell and spatial data is flexible and shared openly, via the MuVIcell Python package, to allow its reuse to profile other cancer types and pathologies.

## 2. Results

### 2.1 Multiplexed ion beam imaging charts the composition and metabolic regulome in the colorectal carcinoma tumor microenvironment.

Healthy human intestinal function relies on a precise interplay of multiple stromal, epithelial and immune cell types[29]. Malignancy disrupts some of these patterns, with metabolic rewiring and spatial reorganization occurring in the TME[5,8–10]. To study the composition and multicellular coordination of colorectal tumor microenvironments, we assembled a cohort of 522 CRC patients and 20 control samples (Fig. 1a). We compiled the corresponding resections on a tissue microarray (TMA)[30]. We designed an antibody panel to segment individual cells, capture lineage, phenotype, and metabolic state (see Methods). This enabled us to locate and quantify corresponding proteins and native metallic elements using MIBI-TOF[28]. In brief, this technology relies on tagging molecules of interest using metal-conjugated antibodies, rastering the sample with an ion beam, and measuring the emission of secondary ions using time-of-flight mass spectrometry. After excluding damaged or depleted cores, we acquired 42-plex images for 470 tumor and 13 control samples. We acquired one image of 400x400 µm (resolution of 390 nm per pixel) per TMA core, each corresponding to a different sample. Curation of these images (see Methods) resulted in 458 tumor cores and 11 control cores with sufficient cells to be analysed.

These samples covered all stages of the disease course as well as different genomic CRC subclasses (Fig. 1a). More advanced node infiltration was observed in later tumor stages (pTNM staging). Patients also varied on the basis of demographics and anatomical location of their tumors (Supp. Fig. S1). Information compiled by trained pathologists and available for most patients included microsatellite instability status, age, sex, tumor size, side and location, RAS mutational status and HER2 amplification status.

To profile individual cells, we implemented a computational workflow (Fig. 1b) to preprocess images, segment cells, and annotate their lineages. Cellular measurements were then stratified to perform several comparisons based on cell type composition, metabolic states, cell shape and spatial organization. After establishing the relevance of these different aspects of the TME biology along disease stages, we aimed to systematically decompose changes occurring in the cohort across cell types, with the assumption that such changes are highly interdependent.

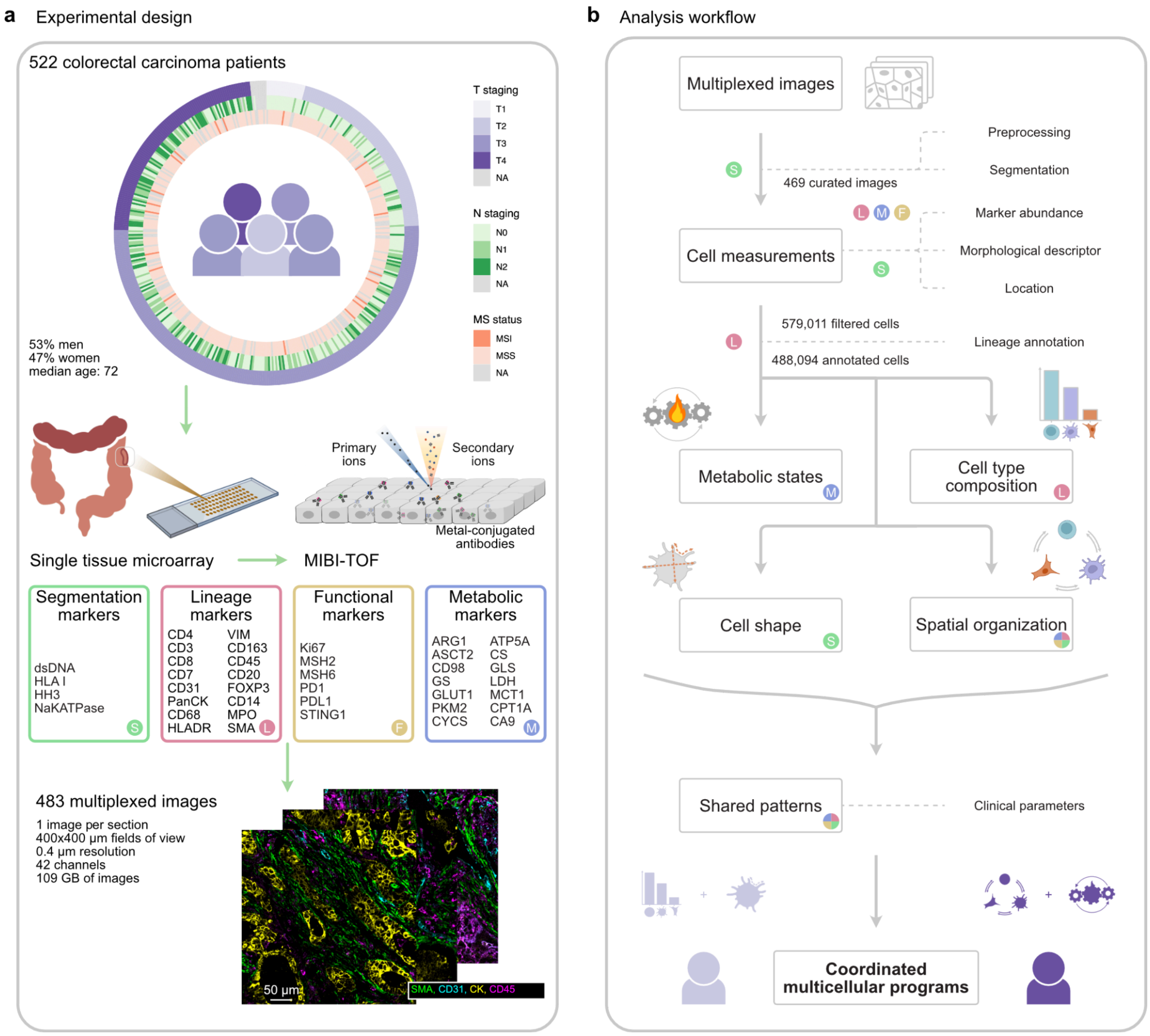

**Fig. 1: Single-cell spatial profiling of the tumor microenvironment in colorectal carcinoma patients.**

**a** Study design from cohort to multiplexed proteomics image acquisition. **b** Analysis design illustrating how starting from the images, we derived single-cell measurements and cell type annotation to perform multiple comparisons between samples, aiming to identify specific multicellular programs underlying disease stages. MS: microsatellite. MSI: MS instability. MSS: MS stability. NA: not available. PanCK: pan-cytokeratin. HH3: Histone H3. dsDNA: double-stranded DNA.

### 2.2 Disease-specific mixtures of cell types compose the tumor microenvironment

We devised a computational pipeline to segment individual cells using a fine-tuned deep learning model[31] (Fig. 2a), while accounting for known technological biases and filtering out potential residual artefacts (see Methods, Supp. Fig. S2a). Based on measured abundance of lineage markers per cell, we inferred cell types in the complete dataset (Fig. 2b, Supp. Fig. S2b). The resulting cell populations had a high abundance of their main lineage markers (Fig. 2c, Supp. Fig. S2c), and cell types were clearly distinguishable based on lineage marker abundances (Fig. 2d). Overall, cancer cells were the most abundant in the regions imaged (53.8% of annotated cells, Supp. Fig. S2d). As previously reported, CAFs were the second most abundant cell type (15.6%)[4,10]. Our phenotyping analysis revealed a considerable diversity in cell type composition between samples, ranging from near-pure epithelial populations to major immune infiltration (Fig. 2e). The relative cell type abundances were structured, with more immune cells found in samples with less cancer cells or with more endothelial cells (Spearman correlation coefficients of -0.34 and 0.34, respectively; Supp. Fig. S2e), matching patterns reported in triple-negative breast cancer[32]. Of note, heterogeneity was also observed within each cell type, with lineage markers not simply being present or absent but falling within bounded abundance ranges (Supp. Fig. S2c,f). We observed significant differences in immune composition between samples, with less $CD8^+$ T cells in the tumor than in the control samples (Fig. 2f). This trend also progressed with tumor stage (Fig. 2g), although other changes were more prominent. In particular, we observed a decrease in all profiled lymphocyte populations. In contrast, we saw a significant increase in $CD68^+$ macrophages and other immune cells in later tumor stages. When considering other common stratification levels of tumors, we observed that the decrease in $CD4^+$ T cells and the increase in $CD68^+$ macrophages was also found in patients with greater lymph node infiltration (later N stages), while no significant change was detected based on microsatellite stability (Supp. Fig. S2g,h). Previous studies identified an increase in immune infiltration in patients with MSI, especially for lymphocytes at the tumor front[33], and our cohort also presented a higher ratio of immune to cancer cells in unstable than in stable samples (Supp. Fig. S2i). However the effect size was small and our analysis revealed that the changes are more complex, with different trends for different immune populations.

Despite the presence of trends across disease subgroups, the heterogeneity between samples meant that no single change in cellular composition was sufficient to perfectly stratify the lesions. To systematically query the relevance and conservation of molecular features shifting along cancer progression, we devised a machine-learning-based predictive modelling approach (Fig. 2h, Supp. Fig. S3a). This allows side-by-side comparisons of different feature types for their combined ability to predict tumor stage in unseen samples, while providing insight into which individual features are the most characteristic of each stage. In brief, we divided the cohort in five for a cross-validation scheme stratified by tumor stage, and kept one of the resulting folds as validation holdout. The remaining four folds were used iteratively for robust model selection. We first applied this approach to the cell type composition measured in each sample. Thus, we obtained an ensemble tree model trained with the XGBoost framework that associated cell type composition and tumor stages (Supp. Fig. S3b). This also linked high T cell counts to the earliest tumor stage (Fig. 2i, Supp. Fig. S2j). The model also performed better than random on the validation holdout (Supp. Fig. S3c,d), although the performance was lower than on the data folds used for model selection. This suggests that the relations between cell types and cancer stages are very heterogeneous (thus favoring overfitting in models presented with parts of the data), yet partly preserved across samples.

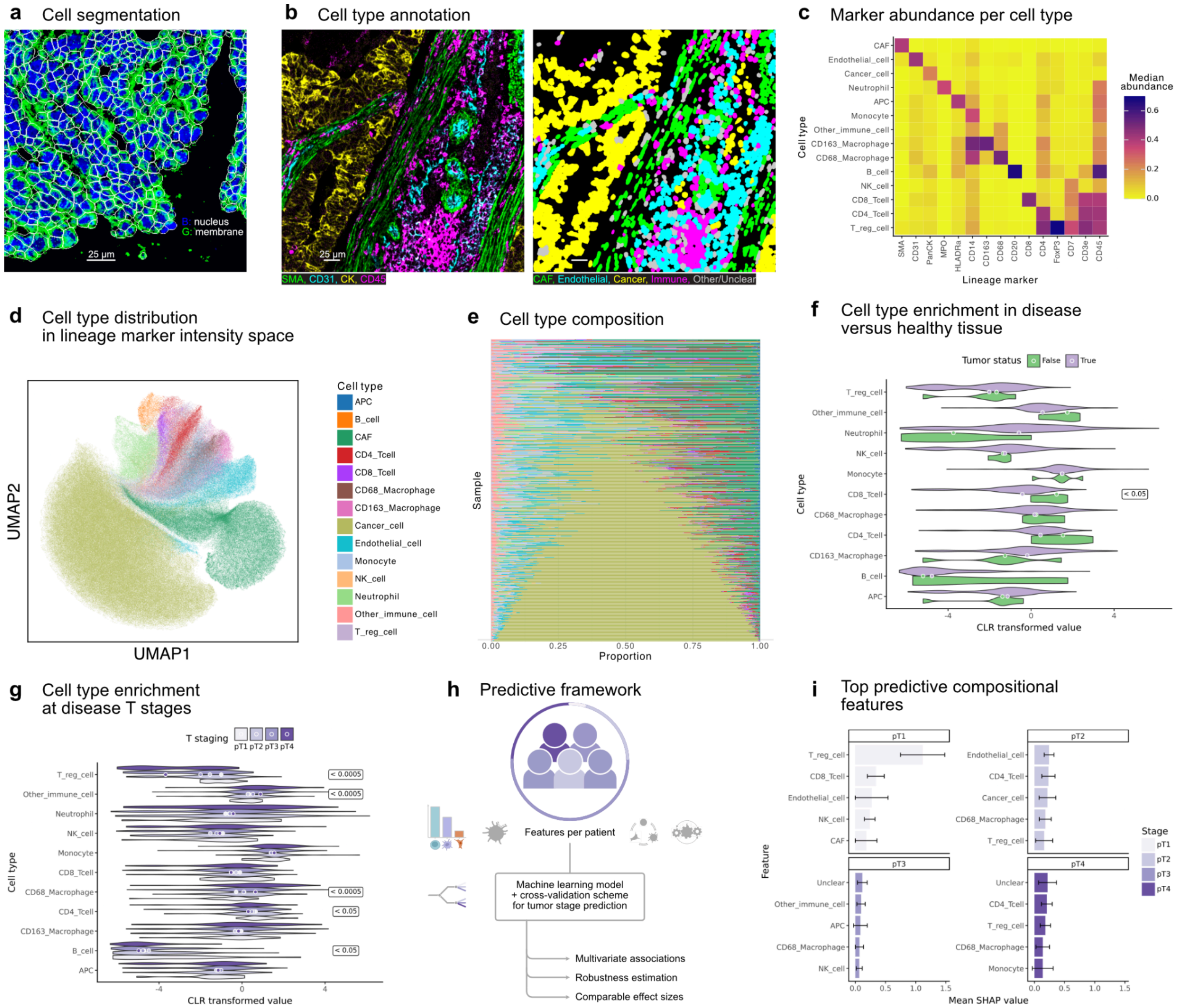

**Fig. 2: Variation in cell type composition across health and disease stages.**

**a** Example of segmented cells after quality filtering in a cropped field of view. **b** Lineage markers and corresponding aggregated cell type annotation for an example field of view. SMA: smooth muscle actin. CK: cytokeratin. **c** Median normalized abundance of lineage markers in the identified cell populations. **d** UMAP representation of all annotated cells based on the intensity of the main lineage markers, displaying the inferred cell type. **e** Proportion of each cell type in each sample, ordered by increasing cancer (epithelial) cell fraction. **f-g** Comparison of cell types present in healthy and tumor samples (f) and at different tumor stages (g). Values are transformed to account for the compositional property of cell type abundances and make them independently comparable, and FDR values < 0.05 are reported for Mann-Whitney U (f) and Kendall's τ tests (g). Central dots represent median values. See Methods for details. CLR: centered log-ratio. **h** Schematic representation of the predictive machine learning framework allowing comparisons of feature associations to disease stage. See Supp. Fig. S3a for details. **i** Mean importance (SHAP values) of top predictive compositional features per stage. Error bars indicate half a standard deviation on each side of the mean.

### 2.3 Adenocarcinoma cells have heightened metabolic capacity

Efforts to stratify CRC patient classes based on molecular data led to the CMS classification, and highlighted common metabolic dysregulations[22]. For instance, CMS3 frequently exhibits a glycolytic switch together with KRAS mutations[23], a combination that has been shown to be therapeutically targetable by glutathione and glucose depletion in lung adenocarcinomas[34]. We further studied the involvement of several metabolic pathways in CRC progression, including glycolysis, fatty acid oxidation (FAO), tricarboxylic acid (TCA) cycle, oxidative phosphorylation (OXPHOS), and amino acid metabolism (Fig. 3a). Additional markers pertained to the proliferation and DNA repair mechanisms in each cell type (MSH2, MSH6, Ki67). Different cell types displayed different metabolic and genomic integrity profiles (Fig. 3b). Macrophages had high amino acid metabolic capacity, in accordance with the established role of arginine in their differentiation and function[35,36]. B cells and cancer cells had the highest proliferation and DNA repair activity. When stratified by stage, cancer cell metabolism became substantially more active than in other cell populations from stage 2 onward (Supp. Fig. S4a). Prompted by the observed relative differences in metabolic potential, we further explored metabolic and functional states specifically in healthy and malignant epithelial cells. Tumor cell metabolism was highly structured, as groups of metabolic markers were consistently high or low together in single cells (Supp. Fig. S4b). We observed a clear set of positively-correlated mitochondrial proteins, an early glycolysis component (GLUT1, PKM2 and LDH), an amino acid transport component (ASCT2, CD98 and GS), and a high correlation between arginine degradation and lactate transport (ARG1 and MCT1). At the patient level, cancer cells showed higher abundance of multiple metabolic proteins (CD98, CytC, MCT1, LDH, GS, GLS, PKM2, GLUT1, ARG1) compared to healthy epithelium (Supp. Fig. S4c). This elevated energy metabolism is consistent with higher energetic needs to support proliferation and malignancy. However, no major change was observed in individual metabolic regulators between tumor stages. With progressing tumor stages, median levels of CPT1A consistently decreased, and levels of GLUT1 increased, yet these changes were smaller than the heterogeneity observed in the abundance of metabolic markers. Because relevant metabolic changes may result from subtle shifts across multiple pathways, we further used our predictive modelling approach on the average and the standard deviation of metabolic and proliferation activity per sample. This model performed better on held-out data (macro-F1 score of 0.293 across stages) than the baseline (macro-F1 = 0.247) and the model based on cell type composition (macro-F1 = 0.273), with no drop in performance compared to the model selection data folds (Supp. Fig. S3b-d). An alternative model based solely on markers of functional states in cancer cells (STING1, PDL1, Ki67, MSH2, and MSH6) achieved slightly worse performance. This points to the relevance of metabolic markers to comprehend disease progression. The correct predictions of the metabolic marker model were largely relying on increased variance per sample in metabolic markers, such as GLUT1 and LDH in pT stages 1 and 4, or CPT1A in stage 2 (Fig. 3c, Supp. Fig. S4d). Inversely, lower spread was observed in PKM2 and ASCT2 in pT3 samples. This highlights the importance of considering the heterogeneity in metabolic marker activities, with a potential for multiple metabolic states, rather than considering average values only. Moreover, higher PKM2 abundance values supported stage 4 predictions, while high Ki67 abundance was often predictive of stage 2 samples. This role of proliferation is consistent with the increases in Ki67, MSH2 and MSH6 observed in tumors (Supp. Fig. S4c). This increase in proliferation, also indicated by a larger fraction of epithelial cells with high expression of Ki67, was stronger in early tumor stages and peaking at stage 2 (Fig. 3d-e). As proliferation could explain a large part of the changes observed, we further disentangled its effect from malignant processes. Aggressive cancer cells had higher metabolic and DNA repair activity (Fig. 3f), although many of these changes were also observed in healthy proliferating cells (Fig. 3g). However, larger increases in ARG1, CA9, CD98, CPT1A, CytC, GS and MCT1 were cancer specific. Taken together, this confirms that some of the metabolic changes observed in tumors are sustaining proliferation, while others are supporting tumorigenic processes in other ways, supporting adaptation to hypoxic TMEs. Of note, only a multiplexed single-cell approach could distinguish between these effects and reveal that metabolic changes in malignant cells serve different purposes.

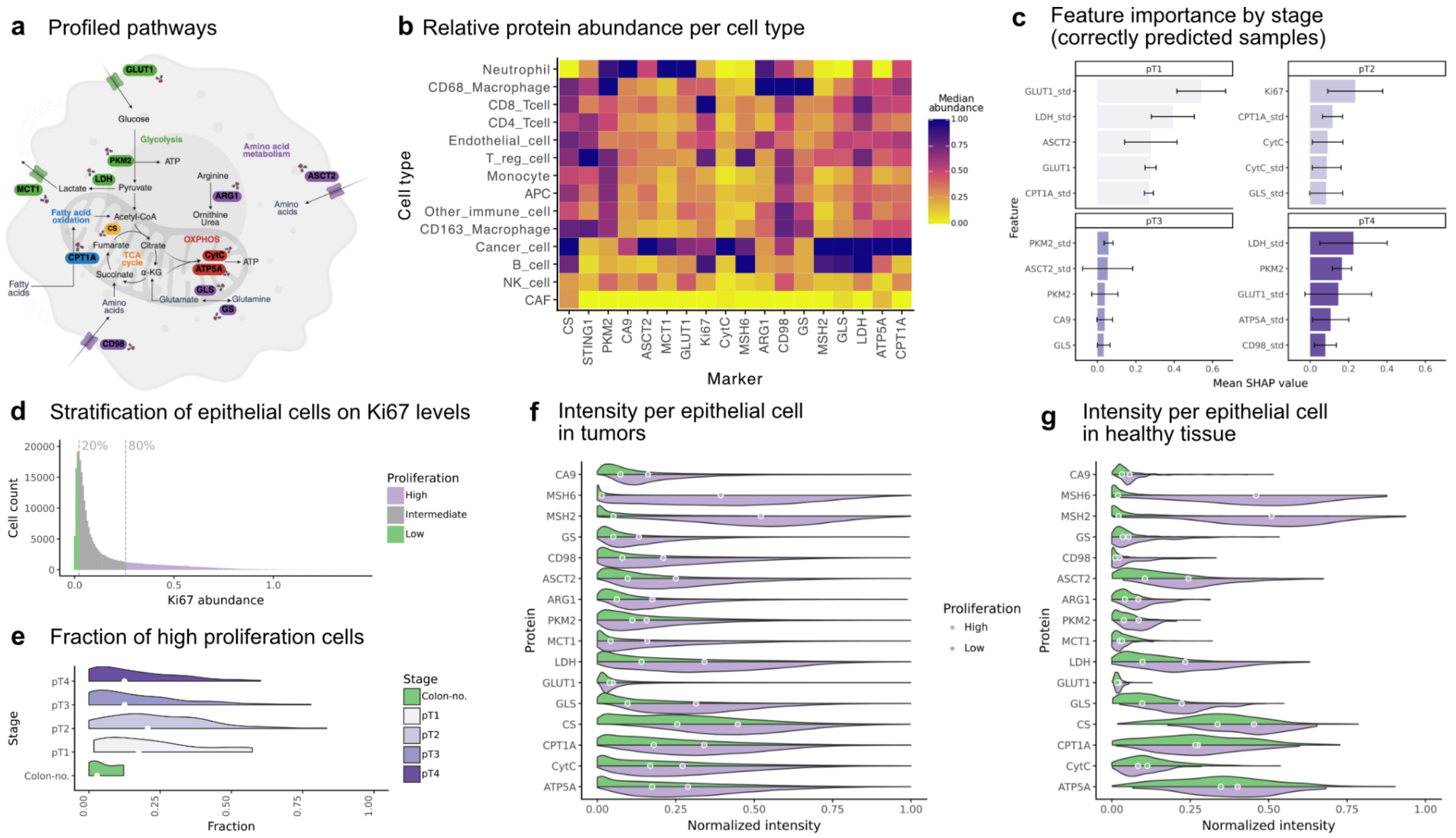

**Fig. 3: Conserved single-cell metabolic signatures in health and disease.**
**a** Diagram of the proteins targeted in the MIBI experiment involved in cellular metabolism. **b** Median metabolic and functional marker abundance per cell type, normalized within each marker. **c** Mean importance (SHAP values) of top predictive metabolic features per stage. Error bars indicate half a standard deviation on each side of the mean. **d** Distribution of Ki67 intensity per epithelial cell, defining high (top quintile) and low (bottom quintile) proliferation populations. **e** Fraction of high proliferation epithelial cells per sample at different disease stages. **f-g** Normalized intensity per metabolic marker in high and low proliferation epithelial cells in tumors (f) and healthy tissue (g). Central dots represent median values.

## 2.4 Distinct spatial patterns underlie disease progression

We further leveraged the spatial information contained in the multiplexed images to explore their complementarity to the cell-level marker intensities. We first studied if cancer cell shape changed with disease progression, as cell shape emerged as a potential source of complementary information in spatial proteomics[37,38]. Cellular morphology, measured by the mean and standard deviation of complementary cell shape metrics per sample, also had predictive value to characterize tumor stages (Supp. Fig. S3b-d). We observed a higher contribution of tumor cells with high eccentricity for pT stage 1, while pT stage 3 stood out by a higher reliance on shape (higher eccentricity) than size (Supp. Fig. S5a-b). The fourth pT stage corresponded to larger cells with little variation in surface area, while stage 2 corresponded to smaller cells (Supp. Fig. S5c).

The tumor and its periphery have a highly structured spatial organization[10]. In stage 3 tumors, homotypic cellular interactions and more frequent infiltrations of immune cells to stromal than epithelial layers have been reported[4]. Thus, we aimed to account for differential rearrangement in space to complement the previous cell type composition analysis we performed. We used MISTy for its ability to identify immune and cellular patterns within the TME, by quantifying the predictability of each cell's lineage based on the identity of surrounding cells[39]. We separated local (juxtaview, neighbors up to 40 pixels or 15.6 µm) and distant (paraview, neighbors between 40 and 120 pixels or 46.9 µm) interaction patterns. This highlighted the high level of spatial organization in the colon in both health and disease, with distinct prominent patterns (Fig. 4a-d). Epithelial cells, whether normal or malignant, were the best predicted cell type from their spatial context (Fig. 4a,c). This is coherent with their abundance and their typical lack of motility. Similarly, fibroblasts and endothelial cells, as other structural cells, could be modelled well. Most of the associations between cell types were homotypic, both in the juxtaview (Fig. 4b,d) and the paraview (Supp. Fig. S5d-e). These patterns corresponded to local clusters of individual cell types, which can often be explained by cell division or by cell-type-specific attraction. Moreover, local neutrophil and $CD163^+$ macrophage interactions and more distant fibroblast interactions were only significant in tumor samples. On the contrary, local APC and B cell patterns were noticed in healthy samples only. This also came with a specific interaction in neighborhoods including $CD4^+$ T cells and B cells but no epithelial cells (Fig. 4e). B cell identities were also particularly well modelled in healthy samples ($R^2$ = 0.501, Fig. 4a). We previously did not observe significant changes in the total abundance of B cells (Fig. 2f), yet B cells were structured in compartments in healthy tissue and less commonly in tumors, even when B cells were abundant in the TME (Fig. 2d,e). There appears to be a shift in the gut-associated lymphatic tissue, from organized structures to diffuse lymphocytes.

We then assessed if the way lineages are spatially associated is also varying across tumor stages using our predictive modelling approach. We found that spatial cell type association was outperforming all other assessed features in predicting tumor stages, both during model selection and on validation data (Supp. Fig. S3b-d). As the arrangement of cell types was a better predictor than their composition alone, this highlighted a broad spatial reorganization happening during the course of the disease. The model relied on several stage-specific interaction patterns that were most commonly homotypic (Supp. Fig. S5f-g). For instance, tight clusters of monocytes were underlying pT1 and pT4 predictions, while $CD8^+$ T cell clusters were positive predictors of pT3 and negative predictors of pT4. Diffuse enrichments of fibroblasts were leading to pT2 but not to pT3 predictions. Some informative interactions also involved multiple cell types, such as enrichment of $CD163^+$ macrophages in the vicinity of $CD68^+$ macrophages or monocytes found as predictive of pT1 and pT2 samples, respectively. CAFs also had a propensity to be found directly next to other CAFs rather than to cancer cells (Fig. 4h), and this pattern was typically weak for pT2 samples and strong for pT3 samples (Supp. Fig. S5g). Indeed, we observed a clearer predictive relationship in later tumor stages, driven by an apparent decrease in immune infiltration of the stroma (Fig. 4i). This highlighted the presence of late-stage tumors with no immune infiltration and mutually exclusive patches of fibroblasts and cancer cells. This could illustrate how late-stage tumors are associated with CAF phenotypes supporting ECM formation and shielding the tumor from immune cells[40]. Similar patterns of lymphocytes sequestered to the stroma were also reported in CRC tumors with chromosomal instability[8]. Overall, this paints the picture of a well-orchestrated spatial organization of tumor, stromal and immune cells shifting during disease progression.

In addition, multicellular metabolic patterns were previously observed in the colorectal TME based on transcriptomics changes, with notable variations in glycolysis across patient groups[9]. Metabolic association between malignant cells and surrounding $CD8^+$ T cells were also identified at the tumor border[11]. This prompted us to also extend our analysis to the spatial dependencies between metabolic markers. Aggregated per sample, the spatial metabolic associations were also predictive of tumor stages (Supp. Fig. S3b-d). These features delivered a good performance during model selection, comparable to the results observed with the spatial organization of cell types, and highlighting the richness of spatially-informed multicellular features. However, spatial metabolic patterns did not perform as well on the holdout as spatial lineage features, which may be more robust predictors by benefiting from prior information on which markers and populations are consistently relevant.

To further study the tendency of metabolic regulation patterns across all cells to be local or global, we leveraged the Kasumi approach which differentiates between recurrent diffuse or localized spatial patterns[41]. In the complete cohort, abundance of multiple metabolic markers could be largely inferred from the abundance of other markers in the same cell, suggesting co-regulation of metabolic enzyme expression (Fig. 4j). Within cells, we saw co-regulated levels of mitochondrial enzymes, and predictive power of lactate transport (MCT1) on amino acid metabolism (Fig. 4k). Furthermore, the abundance of GLUT1, CA9 and Ki67 could be refined using the metabolic levels in surrounding cells (Fig. 4l), highlighting that additional spatial features impact cellular metabolism. Several of these intra- and intercellular patterns were conserved across samples. We also observed positive spatial associations in Ki67 or in GLUT1 (Fig. 4m). This revealed a high level of metabolic organization across cell types in colorectal tissues. Beyond these cohort-wide trends, we could identify motifs localized to restricted tissue regions, yet shared across multiple samples (Supp. Fig. S5h). These motifs, represented by clusters of spatial interactions observed in parts of the images (see Methods), were not uniformly distributed but instead preferentially found in a limited number of samples. The most abundant clusters, such as cluster 7, corresponded to homotypic metabolic relations present in large portions of the tumor and its microenvironment (Fig. 4n). This pattern, as most others, did not show a clear association with disease stage (Supp. Fig. S5i), showing that spatial metabolic regulation is largely conserved during tumor progression, in a similar way to our observations of metabolic states in tumor cells (Supp. Fig. S4c). Another pattern, identified as cluster 11, also shared between samples but less frequent (Supp. Fig. S5h), had similar interactions and was further completed by predictable levels of CA9 based on surrounding abundance of both GLUT1 and CA9 (Fig. 4o). Cluster 11 was specific to late-stage tumors, suggesting a different adaptation to hypoxia in late-stage colorectal carcinoma (Supp. Fig. S5i).

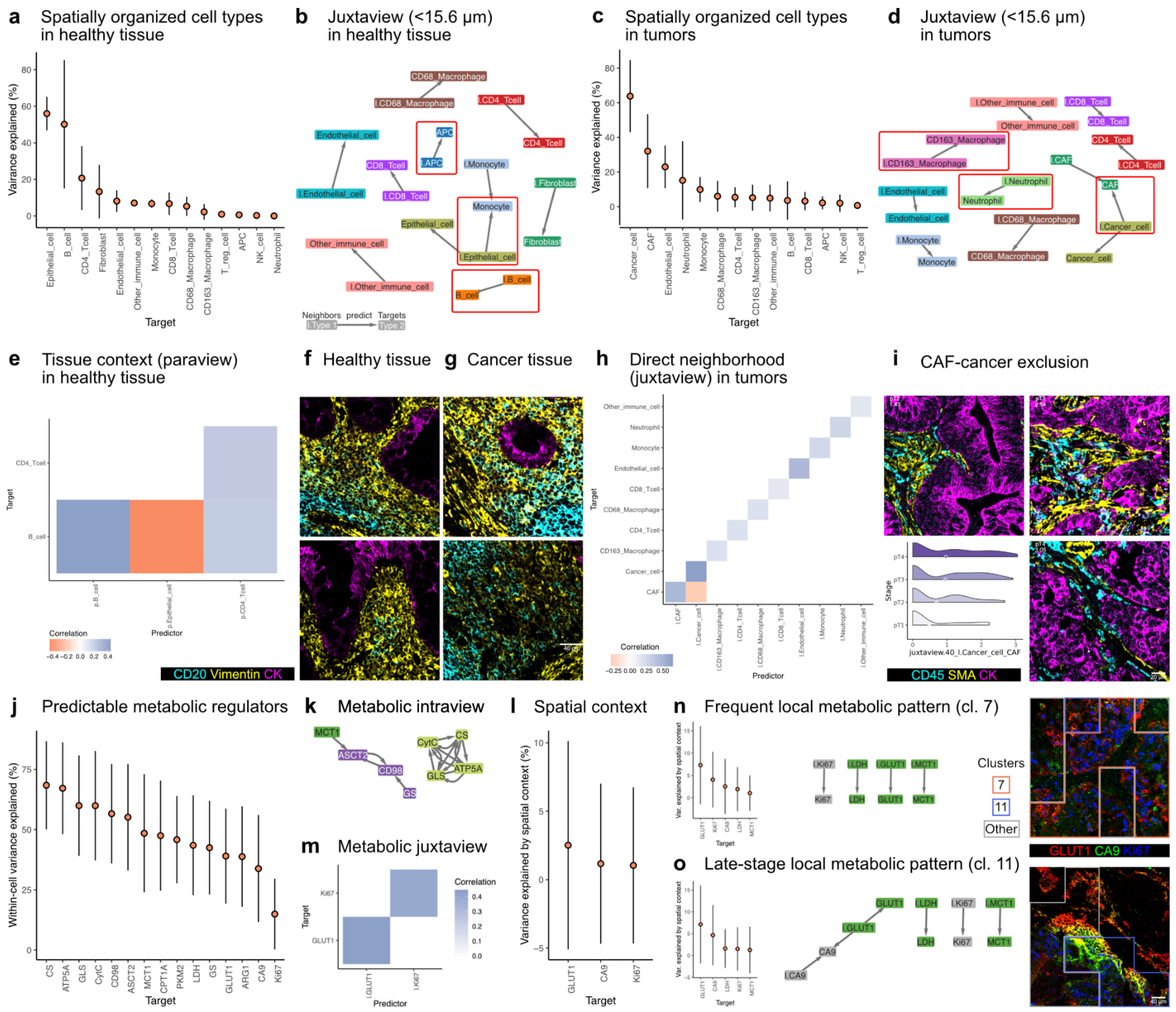

**Fig. 4: Spatial organization into cellular and metabolic niches.**

**a-d** Analysis of the spatial organization of cell types in each sample in healthy tissue (a-b) tumor (c-d). Performance of the predictions per cell type, where points and bars represent the mean and standard deviation, respectively (a,c) and corresponding network of cell type associations in local neighborhoods (b,d). Pairs found only in health or in disease are circled in red. **e** Heatmap of correlation for associated cell types in the healthy paraview. **f-g**

Example images displaying CD20 (B cell marker), CK (epithelial/cancer marker) and Vimentin in B-cell-rich healthy (f) and tumor (g) samples. CK: cytokeratin. **h** Heatmap of correlation for associated cell types in the tumor juxtaview. i Example images CD45 (immune marker), SMA (CAF marker) and CK at different pT stages, along with the effect size per sample of the negative association between cancer cells and CAFs in the juxtaview. Central dots indicate median values. At later stages, being surrounded by cancer cells more strongly predicts that the central cell is not a CAF. **j** Performance on the prediction of metabolic markers based on other markers in the same cell (left) and performance gain based on spatial information (right). **k** Corresponding network of metabolic marker associations within cells (intraview). **l** Heatmap of correlation coefficients for associated markers in the juxtaview. **m-n** Left to right: Gain in prediction performance from spatial information, corresponding network of cell type associations in local neighborhoods and example images for cluster 7 (m) and 11 (n). Images in (m) and (n) are normalized individually to highlight the co-abundant patterns picked up by Kasumi clusters.

## 2.5 The heterogeneous multicellular organization of the TME can be decomposed in discrete factors capturing disease properties

After mapping the CRC TME components one by one, from cell type composition to cell state and spatial architecture, we can now ask how these facets combine or not to generate the full complexity observed across patients. We pursue our analysis by exploring whether co-occurrence or exclusion patterns happen between these different aspects across lineages, by assembling them into a unified framework optimized for the integration of heterogeneous biological data[42]. We combined spatial, morphological, compositional, metabolic and abundance features per cell type and derived a set of 10 comprehensive factors using the MuVI framework[43] (Fig. 5a). To streamline this approach and allow researchers to apply a similar workflow to other datasets and multicellular disorders, we compiled data processing and visualization methods in MuVIcell, an open-source Python package (see Methods, https://doi.org/10.5281/zenodo.17186801). With this package, one can aggregate single-cell measurements per cell type, then identify *factors*, *i.e.* weighted sums of features, potentially spanning multiple measurement and cell types, and recapitulating the most complementary major axes of variation in the cohort. They can be interpreted as activity scores for *multicellular programs*, explaining the *trajectories* along which the samples are distributed during disease progression.

Here, this highlighted 10 factors with different roles, as they varied in the cell types and type of information they described (Fig. 5b). All factors simultaneously comprised information about multiple cell types, showing that all common alterations in the TME involved multiple lineages. Conversely, each cell type was found to change in multiple ways in the cohort, hence associated with multiple factors. The breadth of the changes occurring per cell type varied, e.g. with Factor 10 accounting for the largest share of variance in fibroblast features (0.14 of a total 0.43), while macrophage information was more evenly distributed across factors. Of all the feature types, the factors recapitulated the metabolic descriptors the best (0.87) and the spatial ones the worst (0.05). This is in accordance with highly correlated metabolic markers within cells (Supp. Fig. S4b) and numerous specific spatial patterns (Supp. Fig. S5f-i), as we previously reported. Moreover, the model's objective function weighs individual features equally, so it is more efficient for the model to focus on a large number of metabolic levels that can be explained simultaneously than on sparse and independent spatial relationships. Multiple factors were also associated with clinical variables, which the model had no information about when decomposing molecular and spatial features. Factors 2 and 6 were associated with pT stage and largely independent of each other (Fig. 5c). They represent distinct tumor progression axes, and individual lesions often aligned with one of them, supporting the presence of coordinated TME organization programs. They share multicellularity and hypoxia in early tumor stages (pT1 and pT2), but reflect different malignant developments.

Factor 2 had its highest values for pT1 and pT2 samples then saw a progressive decline at later stages (Fig. 5d). Thus, in tumor samples, features with positive loadings decreased with stage, while those with negative loadings increased with stage. Higher Factor 2 values corresponded to increased abundance of multiple metabolic and functional markers in most cells. This included a high contribution of fatty acid oxidation (CPT1A) and hypoxia (CA9), with a noticeable covariation of amino acid transport in epithelial and cytotoxic cells. Some observations were also more specific to particular cell types: MCT1 and ARG1 loadings were negative in fibroblasts, which also had neutral or slightly negative loadings for other metabolic markers, while PDL1 was especially high in macrophages. A rise in fibroblast metabolism in late stages suggested a stromal participation in lactate circuits consistent with the reverse Warburg effect. This is also consistent with our observations of spatial organization that revealed that CAF-tumor cell interactions prevail at later stages (Fig. 4i). In these late lesions, we also observed a loss of FAO-rich, PDL1-high M2-like macrophages known to favor tumor growth[44], which is coherent with the lower proliferation observed at later stages (Fig. 3e). This complemented our observation that $CD68^+$ macrophages but not $CD163^+$ M2-like macrophages were more prevalent at later stages (Fig. 2g), and further reinforced that TME development could initially be shaped by immune factors and later by epithelial-stromal interactions. We collectively refer to these changes as the tumor-CAF multicellular program, which more precisely matches a TME development from an initial phase of nutrient-driven competition between tumor and immune cells to a metabolic cooperation between tumor and stromal cells.

Factor 6 progressively decreased with increasing pT stage. This factor had higher values for more epithelial cells and less myeloid cells (Fig. 5e). As such, this matched the presence of tumors with increasing myeloid cell fraction at later stages we observed previously (Fig. 2g). OXPHOS (CytC, ATP5A) and lactate transport (MCT1) showed the strongest metabolic loadings, especially in immune cells. Other positive loadings across lineages included CS, GLS, MSH2 and CA9. GS had negative loadings, strongest in endothelial cells, which contributed less to most parameters. This suggested that early-stage tumors were relying on OXPHOS and mitochondrial metabolism, while a shift towards glycolysis occurred at later stages, which is less efficient but allows such tumors to reach a higher mass despite hypoxia[45]. MCT1 and CA9 activity is detected early, pointing to the interplay between hypoxia and lactate metabolism without strong endothelial activity. Later as factor values fall, rising endothelial GS indicates a metabolic remodelling of the vasculature and the activation of angiogenic processes. This mechanism, through which lactate acts as a pro-angiogenic signal in oxidative cancer cells, is well-described in cancer cells[46] but also appears to be visible at the level of a metabolic niche in the TME. We refer to the evolution of the parameters underlying factor 6 in advancing tumor stages as the glycolytic shift multicellular program. Together, the tumor-CAF and the glycolytic shift programs were largely independent and also revealed common metabolic insufficiencies at later stages in cytotoxic lymphocytes, with a marked reduction in amino acid metabolism, that could explain inferior tumor control.

We also noted other recurrent patterns of multicellular reorganization of the TME, as picked up by our factor analysis, that were linked to clinical features. The lymph node infiltration stage was associated with two additional factors (Supp. Fig. S6a). A first tendency for primary tumors with lymph node metastases was to display higher metabolic capacity in epithelial cells and higher glycolytic activity across cell types. This supports the need for aerobic glycolysis to drive EMT and metastatic competence, established in other cancer types, also in CRC[47]. A second pattern observed in metastatic tumors was lower metabolic capacity in endothelial cells and fibroblasts, with larger sizes for all lineages. On that note, changes in endothelial metabolic subsets are often leading to vessel sprouting defects[48], and vessel normalization and immunomodulation offers potential innovations in cancer therapies. The MS status was also associated with two factors (Supp. Fig. S6b). The first showed reduced lymphocyte fraction but high glycolysis (PKM2, GLUT1) and hypoxia across lineages, together with elevated mitochondrial metabolism in lymphocytes, fibroblasts, endothelial and cancer cells. The second showed high STING1, CS and ATP5A across lineages and low hypoxia (CA9). Trends depicting limited CD4+ T cell infiltration and decreased hypoxia were previously reported in MSI tumors[9], and our analysis shows that these are two distinct mechanisms and do not necessarily co-occur. We also confirm a strong metabolic rewiring coupled with the activation of immune recognition of MSI tumors via STING, which is central to the effective antitumor immune response[49,50]. Such programs highlight that MSI tumors can combine immune exclusion with metabolic stress, or alternatively maintain mitochondrial competence and STING-driven immunity. Altogether, this showed that the multicellular remodelling of the TME is not only informative of direct properties of the primary tumors, but can also form a more comprehensive description of the type and state of the disease.

To better understand the coordination within the corresponding multicellular programs, we quantified how well pT-associated features could be predicted across cell types. We focused on the five features with the highest absolute loadings in Factor 2 or 6 per cell type (Supp. Table 1). Then, we optimized classifiers in a cross-validation scheme to predict these features across cell types. Features of one cell type that strongly predicted features of another were taken as evidence of coordination. Since these features define the major axes of variation linked to tumor stage, the resulting predictions can be interpreted as a map of multicellular events during tumor progression. Regulatory relations identified in early-stage tumors linked all immune cells, with weaker ties to endothelial cells (Fig. 5f). The same cell types were connected in late-stage tumors, and fibroblast features also predicted those of endothelial cells and $CD4^+$ lymphocytes (Fig. 5g). Macrophage features were also more predictable, to a large extent from $CD4^+$ T cell features, highlighting very organized myeloid phenotypes in advanced lesions. Of note, in these cell type influence networks, all coarse cell types studied were interconnected, with the exception of cancer cells. This emphasizes coordinated multicellular changes in the microenvironment along disease trajectories involving both stromal and immune compartments, but largely independent of the state of malignant cells. Underlying the accurate cross-cell-type predictions, several functional and metabolic markers showed high correlation in their abundances across cell types within each sample (Fig. 5h). This further emphasizes that the TME structures itself into specific metabolic and functional niches where different cell types vary in a coordinated manner. The correlation patterns also displayed more similarity between similar markers across cell types than between different features of the same cell type. Thus, we further looked into the complementarity of the different types of descriptors derived from our spatial readout (abundance, metabolic and functional markers, spatial organization and cell shape). As for the cell-type-centric influence network, we selected the top features contributing the most to pT-associated factors and performed cross-feature-type predictions. No pair of feature types led to predictions with $R^2 > 0.2$, showing the high complementarity of the different scales and information layers to accurately describe changes occurring during the course of the disease.

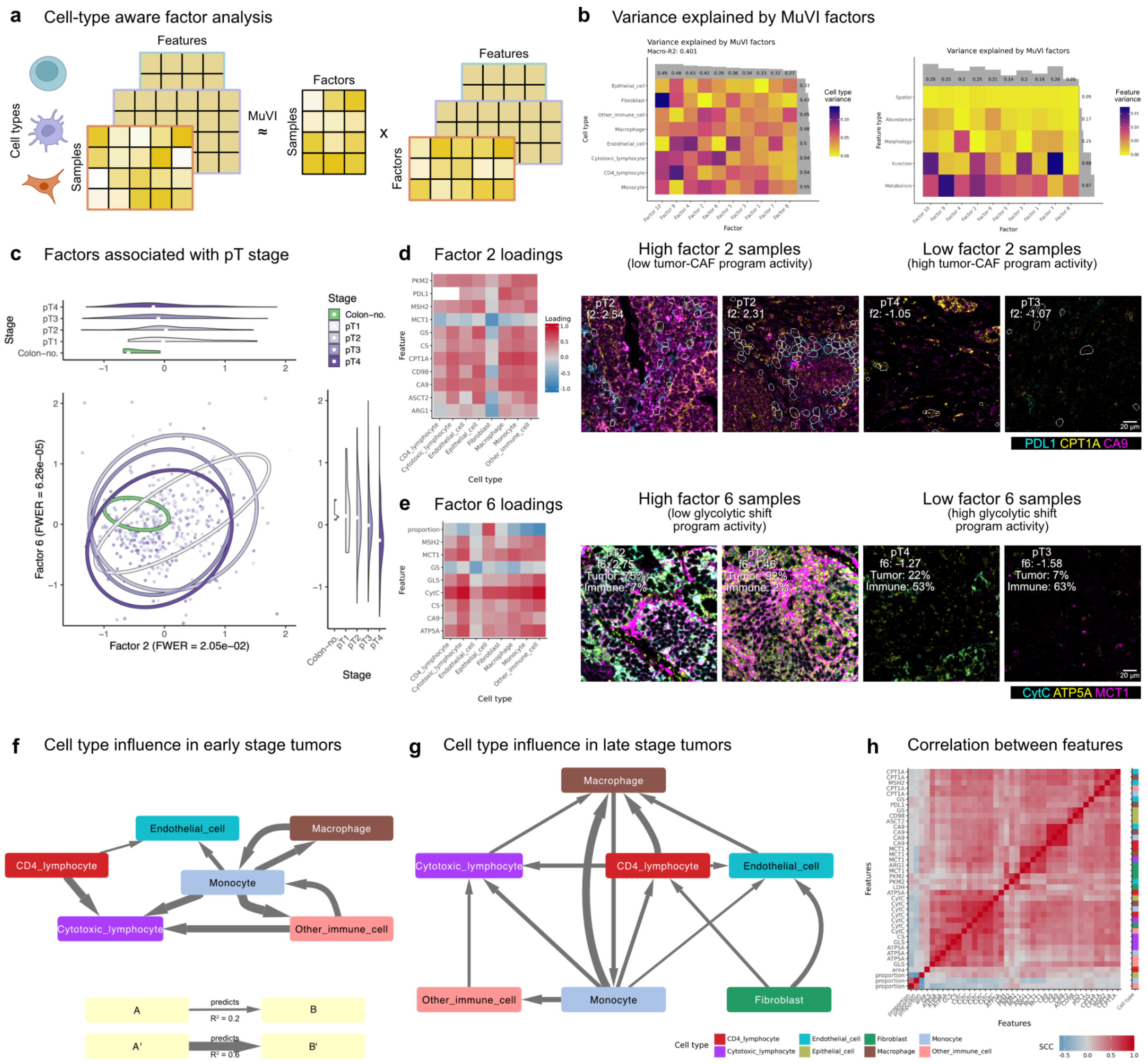

**Fig. 5: Coordinated multicellular and multimodal programs capturing TME changes in the complete cohort.**

**a** Diagram of the factor analysis stratified by cell type and including all major types of features (abundance, function, metabolism, morphology, spatial) previously identified as disease relevant. **b** Fraction of variance explained ($R^2$) by each factor when reconstructing features for a given cell type (left) or per feature type averaged across cell types (right). **c** Multicellular factors associated with pT stage. Family-wise error rate (FWER) reported for Kruskal-Wallis rank sum tests, corrected to account for the association being tested for all 10 factors. Confidence ellipses show the region within 2 standard deviations from the mean per clinical variable level. Ranges are selected to best show the ellipses, with outlier samples truncated. Central dots represent median values. **d-e** Loadings of top features per factor and example images with extreme factor values. In (d), stage and Factor 2 (f2) values are reported, and macrophages are outlined in white. In (e), stage, Factor 6 (*f6*), fraction of tumor (*Tumor*) and immune cells (*Immune*) are reported. **f-g** Networks showing which of the top 5 features underlying pT-associated factors for a cell type are predictive (outgoing edge) of the top features for another cell type, when validated on pT1 and pT2 (f) or pT3 and pT4 (g) samples. **h** Spearman correlation coefficient (SCC) between selected features across samples.

## 2.6 - Multicellular coordination and stage-specific patterns are conserved and confirmed across patient cohorts

Next, we aimed to confirm and further characterize our findings in independent cohorts. First, we reanalysed single-cell transcriptomics data from a study from Joanito and colleagues[51]. The authors identified cell types in the tumors of 63 patients with known pT stage, allowing a direct comparison to the TME composition we inferred from MIBI images (Fig. 6a). In combination with the high inter-tumor heterogeneity, the number of patients limited the statistical power of the analysis. Yet, similar trends were observed as in the spatial proteomics approach, with a decrease in the abundance of all lymphocytes at later stages. The larger number of markers available also permitted defining different cell type populations, suggesting a shift from plasmacytoid dendritic cells (pDC) to merocytic dendritic cells (McDC) in later stages.

Next, we assessed whether the differences in metabolic regulators observed in aggressive malignant cells were present at the transcript level, based on a cohort of 64 CRC resections and 36 adjacent normal tissues, profiled using single-cell sequencing by Pelka, Hofree, Chen and colleagues[9]. Regressing out the effect of proliferation by normalizing expression against *MKI67* counts, we observed more activity in malignant epithelial cells of the metabolic regulators previously identified as cancer specific (Fig. 6b). The strongest changes occurred for *CD98* and *CYCS*. The broad coverage of scRNA-seq allowed us to further compare proliferative cancer cells to healthy proliferative epithelial cells and look into the processes specific to malignancy. This revealed major expression changes (Fig. 6c).

In healthy proliferative cells, this included higher expression of carbonic anhydrase genes (log2-fold change >= 5 for *CA1*, *CA4* and *CA7*) and other established CRC suppressors such as *SPIB* and *GUCA2B*[52]. In aggressive tumor cells, the top overexpressed genes corresponded to known CRC markers and investigated targets such as *NOTUM*[53], *REG3A*[54] and *IGF2*[55]. Looking at the whole ranked list of expression changes, we identified patterns matching higher activity of MAPK, EGFR, JAK-STAT, PI3K, VEGF, NF-κB, TNFα, WNT, estrogen and TGFβ signaling pathways and a lower p53 activity (Fig. 6d), which strongly aligns with the pathways known to be hijacked in CRC cells[56]. These changes could be attributed to a higher activity of a limited set of transcription factors, with strong enrichment for *CTNNB1*, *AP1*, *ISL1*, *TCF7L2* and *ARX* which regulated a network of signaling molecules and other transcription factors (Supp. Fig. S7a). Thus, the structured molecular changes in tumor cells can be attributed to a limited set of transcription factors and pathways, driven by intrinsic perturbations and extrinsic TME signals.

To validate and complement the spatial organization patterns we identified, we used spot-based spatial transcriptomics experiments conducted by Heiser and colleagues[8]. This included 47 samples from 29 patients with known tumor grade. In brief, the authors inferred lineage signatures telling about the cell types present in each sequenced spot, and we derived additional features based on PROGENy signaling footprints[57] and MSigDB metabolic pathways[58]. First, we saw that increased activities in MAPK, EGFR, JAK-STAT, PI3K, VEGF, TNFα and TGFβ, and decreased activity in p53 signaling were also visible across all spots and became stronger at later tumor grades (Supp. Fig. S7b). Using cell type transcriptomic signatures to proxy for composition per spot, we observed that both fibroblast-epithelial and fibroblast-immune exclusion were stronger in late-grade patients (Fig. 6e, Supp. Fig. S7c). This reinforces the notion that CAFs tend to shield advanced tumors (Supp. Fig. S7d). We also confirmed an increased association between *CA9* and *SLC2A1* (coding for GLUT1) per spot in more advanced tumors (Fig. 6f), matching the emergence of the spatial interaction cluster we previously identified (Fig. 4o).

We then studied the relative reliance on key energy metabolic pathways. To do so, we scored if the deviation to the median metabolic expression profile per sample could be attributed to increased glycolysis, OXPHOS or FAO levels (see Methods). We saw different combinations of all three metabolic profiles, and a visible absence of samples entirely dropping usage of a pathway (p-value < 1e-5) (Fig. 6g). We found that early-grade tumors were enriched in relative FAO usage (mean G1 distance to FAO vertex: 0.513, empirical p-value = 0.0061), in accordance with Factor 2 we previously derived (Fig. 5d). We also observed that OXPHOS-favoring tumors were richer in epithelial content and lower in immune content (Fig. 6h), corresponding to the behavior described by Factor 6 (Fig. 5e). To quantify the association between molecular features and tumor progression in this validation cohort, we trained classifier models to take sample-level information on lineage signature activities and predict the corresponding tumor grade. We applied this approach to different feature sets: the epithelial signatures, the immune and stromal signatures, and a combination of all of them. As in our prior modelling efforts based on spatial proteomics data, the TME features were informative of the tumor state and all models performed significantly better than random (Fig. 6i). Here, TME information also complemented the tumor-intrinsic information, with the models using the complete set of lineage signatures performing the best.

Exact coordination patterns could not be resolved in previous studies due to volume and technological limitations. Yet, individual aspects from the multicellular programs we identified were consistently supported in independent patient cohorts. This added evidence and context to the distinct *archetypes* of TME organization observable in advanced tumors (Fig. 6j), representing the extreme configurations reached along the major malignant trajectories. In early tumors, malignant cells were often more proliferative and mitochondrially active (FAO, OXPHOS). In late-stage tumors, a first archetype was characterized by higher glycolysis, vasculature remodelling and immune compositional changes, including more myeloid cells. Another archetype corresponded to prominent tumor-stromal interactions, with clear separation between tumor core, stroma and immune cells.

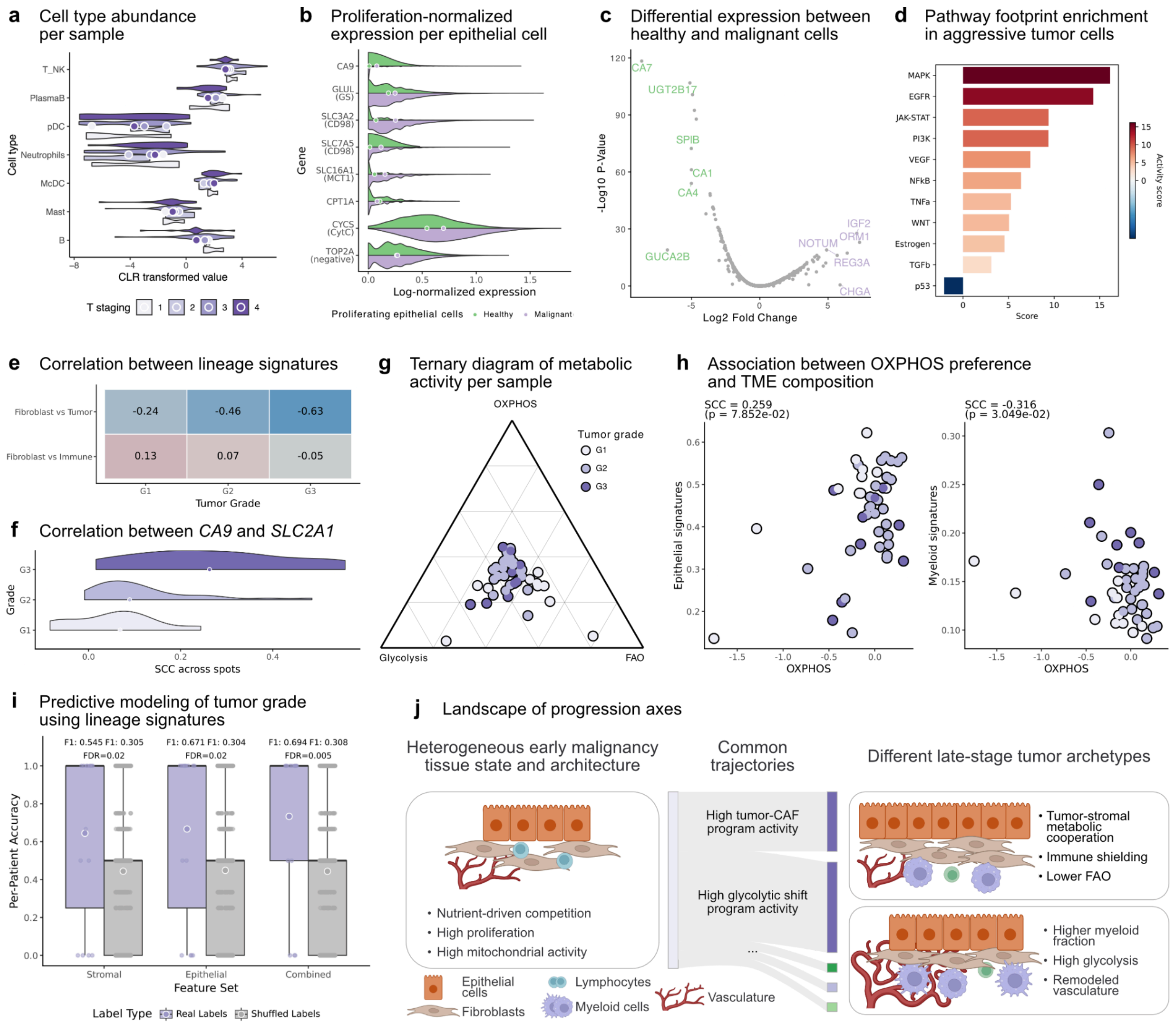

**Fig. 6: Multicellular programs conservation and transcriptional regulation.**
**a** Centered log-ratios (CLR) of the cell type proportions stratified by stage. **b** Metabolic gene expression regressed for cell proliferation in proliferative malignant and healthy epithelial cells. TOP2A is added to show that proliferation-related processes are efficiently regressed out. Central dots represent mean values. **c** Differential gene expression between proliferative malignant and healthy epithelial cells. **d** Corresponding enriched pathway footprints. **e** Spearman correlation coefficient (SCC) between fibroblast and tumor (upper) or immune signatures (lower) per spot. **f** Correlation between spot-level expression of CA9 and SLC2A1 (coding for GLUT1), stratified by tumor grade. **g** Ternary diagram showing the relative fractions of normalized deviation of Glycolysis, OXPHOS and FAO signatures.
**h** Relation between median OXPHOS and epithelial or myeloid signatures per sample.
**i** Predictive performance in tumor grade prediction based on different sets of lineage signatures. Accuracy in multi-sample patients (n = 15). Labels were shuffled 100 times. False discovery rate (FDR) obtained using the Benjamini-Hochberg method on the p-values of Mann-Whitney U tests comparing the distribution of the accuracy values for predictions of the real and shuffled labels. Scores reported are macro-F1 scores across samples. **j** Outline of the multicellular coordination programs identified and conserved across cohorts. Data from Joanito et al[51] in (a). Data from Pelka, Hofree, Chen et al[9] in (b-d). Data from Heiser et al[8] in (e-i).

## 3. Discussion

In this study, we asked how metabolic reprogramming and spatial organization of the TME relate to disease stage across 522 CRC patients. Using high-dimensional multiplexed imaging and a bespoke machine-learning framework, we systematically dissected changes in the TME and identified co-occurring patterns across cell and feature types. This is, to our knowledge, the most extensive single-cell and spatial study of human CRC to date. Latent factor modelling revealed two robust multimodal and multicellular programs corresponding to progression axes. One captured early FAO rewiring with prominent fibroblast-epithelial interactions. The other reflected a TME shift toward glycolysis with likely increased endothelial metabolic activity. CRC progression follows multiple, partly independent trajectories, and not a single path. Our approach also offers a scalable workflow, powered by the MuVIcell package, applicable to other solid tumors.

Lesions were highly heterogeneous in their composition and shifted from lymphoid-rich early-stage tumors to myeloid-rich late-stage tumors. Metabolic rewiring happened as early as stage pT1, with declining CPT1A and rising GLUT1 levels. However, the dominant signal was metabolic heterogeneity rather than a single monotonic trend. These dynamics were guided by multicellular metabolic niches rather than tumor cells in isolation. Spatial features, including homotypic clustering, CAF-cancer cell proximity, and immune cell infiltration, predicted stage more strongly than cellular composition and metabolic descriptors. Non-epithelial compartments better reflected tumor stage than malignant epithelial metabolism alone, highlighting the importance of multicellular interactions. Moreover, we validated our results using independent single-cell and spatial transcriptomic datasets[8,9,51]. We confirmed stage-linked metabolic axes, changes in immune infiltration, and stromal alterations. This provided additional evidence for the different trajectories between stages, and positioned multicellular spatial organization and metabolic niches as key markers of disease progression.

Metabolic reprogramming, especially glycolytic shifts and hypoxia adaptation, is an established hallmark of CRC[5,7,59]. Emerging spatial proteomics and transcriptomics techniques also evidenced spatial context and multicellular coordination as central to colorectal tumors[1,6,8–10]. We extend these findings by providing quantitative insight into the spatial and metabolic changes and their coordination. Thus, we found that their heterogeneity results from intrinsic variation, intercellular effects and differential intra- and intertumor progression trajectories. Metabolic states were also spatially organized into niches encompassing epithelial, stromal and immune cells. We extracted the most prominent axes of variation in the cohort by deriving 10 factors, aligned in part with hubs identified in transcriptomics studies[9]. The factors aligned with known processes, such as CAF mediation of tumor growth and immune interactions[12]. In addition, we contextualized these findings, for instance by describing co-occurring glycolytic shifts and malignant angiogenic processes, with higher myeloid fraction. This shows the relevance of multicellular metabolic and spatial changes in combination, as a therapeutic approach would have to act within this remodelled TME. In this case, treatments targeting glycolysis in cancer cells, a strategy with growing interest[60,61], would occur in a different immune context and would likely impact immune phenotypes as well. Impairing glycolysis in the abundant myeloid population could steer them towards immunosuppressive phenotypes[62], so it could be beneficial to additionally target myeloid cells in the tumor vicinity. Overall, the novelty of our findings stems from combining state-of-the-art machine-learning with robust spatial single-cell proteome quantification. This integration provided precise functional and spatial insight that could not be obtained with spot-based spatial transcriptomics, as it required directly observing the local arrangement of single cells. This further motivates the development of refined analytical and technological tools to elucidate the mechanisms underlying TME reorganization and guide therapeutic advances.

Some limitations can be noted. Antibody panels limited pathway coverage and prevented direct metabolic flux quantification. We addressed this by complementing protein abundance with broad transcript-based signatures and by focusing on rate-limiting enzymes and key transporters, respectively. Follow-up assays could add functional layers by measuring metabolites, integrating microbiome profiles, tracking tumor and lymphocyte clones, or quantifying antigen-presentation and cell-cell signaling. We focused on pT, extensively annotated in the cohort and among the strongest predictors of patient survival[3,25,26]. We were also able to identify major multicellular changes associated with metastatic potential and genomic classes. However, we lacked direct survival, pre-malignant lesions and post-treatment information. Future work could investigate associated factors and underlying features to serve as clinical markers. This also highlights that the resolution and descriptive power of MIBI are ideal for in-depth cohort characterization, but simpler readouts could be used for clinical applications. For instance, histological staining would suffice for morphology and tissue composition, and immunohistochemistry for individual markers. Additional data could also help refine the number of factors and the granularity relevant for different clinical tasks, or the regulation of the carcinogenic processes we observed. With a suitable methodology, their temporal dynamics, causal relations and functional consequences could also be studied further. Therapeutically, targeting hypoxic vascularization[48,63], immune-metabolic shift to glycolysis[7], and CAF activity and metabolism[12,64] are promising avenues that align with the programs we observed, and data-driven polypharmacology could be considered.

Finally, we share spatial data, derived features, and open-source and flexible code to reproduce all analyses (https://doi.org/10.5281/zenodo.17008987) and apply it to other cell type-stratified measurements (https://doi.org/10.5281/zenodo.17186801). As such, our approach can easily be extended to other solid tumor types to guide the discovery of biomarkers and therapeutic targets.

In summary, CRC progression is shaped by coordinated shifts in metabolic niches and spatial organization. The multicellular programs we identified provide new perspectives on TME organization in CRC and may help further stratify patients based on TME architecture and tumor cell states.

## 4. Methods

### Clinical samples

FFPE samples from 522 CRC resections and 20 control tissues were collected retrospectively at the Institute of Pathology of the University Medical Center Hamburg-Eppendorf, and compiled on a single TMA.

After image acquisition and data curation, we were left with a subset of 458 patients, for which the main available clinical features are visualized in Supp. Fig. S1 and the cohort annotation is summarized in Supp. Table 2.

### Antibody panel preparation and validation

To ensure reproducibility and robustness of our panel, we thoroughly tested all antibodies used in this study for antigen specificity and selectivity. To this end, we performed immunohistochemistry on human tissue microarrays containing both healthy and malignant tissues, adhering to established protocols[65,66]. In short, FFPE samples were sectioned at 5µm tissue thickness, baked for 30 minutes at 70°C and rehydrated using a reverse ethanol series. Antigen retrieval was conducted at pH9 and 97°C for 40 minutes. Endogenous peroxidase was inactivated using 3% H2O2 for 30 minutes. Off-target binding of antibodies was blocked using a protein-rich blocking buffer for 1h at room temperature and followed with primary antibody incubation (diluted in 3% normal horse serum) at 4°C overnight. Both of these steps were performed using the Epredia Sequenza staining platform (Thermo Fisher Scientific). On the next day, chromogenic development was performed by adding 3,3'-diaminobenzidine (DAB, VectorLabs) for 40 seconds. Nuclei were counterstained using Harris' hematoxylin solution (Sigma) with an incubation time of 10 seconds. Subsequently, slides were dehydrated using a progressive ethanol series and mounted using the VectaMount permanent mounting medium (VectorLabs). We assessed antibody specificity by visually comparing the observed spatial signal distribution across different healthy and diseased tissues, as well as signal-noise ratios, to reference images from Human Protein Atlas[67]. Selected clones were subsequently conjugated to heavy metal reporter tags using antibody conjugation kits (Ionpath) according to manufacturers instructions. The final panel, including staining concentrations in MIBI, is listed in Supp. Table 3.

**MIBI staining and acquisition**

Tissue sections (4 µm thick) were cut from TMA FFPE tissue blocks using a microtome and mounted on gold-coated slides (Ionpath) for MIBI analysis. Slides were incubated at 60 °C for 1 h in a drying oven to facilitate paraffin melting. Deparaffinization was carried out using three consecutive washes in xylene, followed by a graded ethanol rehydration series (2x 100%, 2x 95%, 1x 80%, 1x 70%) and two final rinses in ultrapure water (Millipore). All steps were automated on a Leica ST4020 Linear Stainer (Leica Biosystems). Antigen retrieval was performed at pH 9 using the Target Retrieval Solution (DAKO, Agilent Technologies) at 97 °C for 40 min, followed by a passive cooling phase to 65 °C over 50 min, conducted on a Lab Vision PT Module (Thermo Fisher Scientific). Sections were subsequently washed twice in PBS supplemented with 0.1% (w/v) bovine serum albumin (BSA; Thermo Fisher Scientific), then blocked for 1 h at room temperature in a blocking buffer containing 2% (v/v) horse serum, 0.1% gelatin (Sigma-Aldrich), 0.1% Triton X-100 and 0.02% sodium azide in TBS IHC Wash Buffer with Tween 20 (Cell Marque). Tissues were incubated overnight at 4 °C with a multiplexed antibody master mix prepared in antibody diluent and filtered through a 0.1 µm polyvinylidene fluoride (PVDF) centrifugal membrane (Millipore). Following incubation, slides were washed twice in the wash buffer and fixed in 2% (v/v) glutaraldehyde (Electron Microscopy Sciences). Slides were washed thrice with 100 mM Tris (pH 8.5) and twice with Millipore water. Dehydration was performed using a graded ethanol series in reverse order (1 × 70%, 1 × 80%, 2 × 95%, 2 × 100%). Then, they were vacuum-dried and stored desiccated until further analysis.

Multiplexed images from these slides were acquired using the MIBIScope (Ionpath, Cat. No. 1) operated through the MIBIcontrol software (v1.8.0). Imaging areas were selected to be 400 x 400 µm fitting into each TMA core. All quality control steps were followed according to the manufacturer's guidelines, and stigmation was automatically performed by the instrument. Images were captured with a dwell time of 1 millisecond in "fine" mode, resulting in image resolutions of 1024 x 1024 pixels.

**Image extraction and preprocessing workflow**

Images were derived from mass spectrometry data using Toffy (https://github.com/angelolab/toffy). For each pixel, the intensity corresponding to a specific mass *m* was calculated by integrating the ion counts over the interval [*amu* – 0.3, *amu*]. The resulting images were subsequently processed using Rosetta compensation to mitigate cross-channel contamination as previously described[68].

The mass and relative composition of these interfering species in the source reagent are well characterized, allowing correction by scaling the source channel intensity prior to subtraction from the target channel.

We observed a consistent spatial gradient across images and most channels, which we identified as an instrumental artifact. To correct for this gradient, we first estimated it by suppressing sample-specific signals. Specifically, we averaged the channels corresponding to markers that are ubiquitously expressed across all fields of view (CD98, CytC, MCT1, ASCT2, LDH, GS, GLS, ATP5A, CS, PKM2, GLUT1, CPT1A, MSH2, dsDNA+HH3). This averaging step minimized signal variance arising from tissue architecture. The resulting average was then strongly blurred and summed to produce an initial estimate of the gradient matrix. This matrix was normalized by its 99.9th percentile to yield a relative gradient profile. We found that the prominence of this gradient was channel-specific and inversely related to the mass of the channel. To account for this, we computed gradient prominence as the intensity difference between the 0.1st and 99.9th percentile pixels for each ubiquitously expressed marker channel. We then performed a simple linear regression of gradient prominence against channel mass, using the regression output to estimate gradient strength for all channels. This approach mitigated bias from incomplete imaging window coverage by markers expressed only in certain subsets of cells. To obtain the corrected images, each channel image was divided by the relative gradient profile scaled according to its regressed prominence.

**Segmentation and cell filtering**

To position individual cells in each image, we first used CLAHE correction[69] on the nuclear marker channel (dsDNA+HH3) and a sum of membrane-displaying channels (NaKATPase + GLUT1 + ASCT2). In the resulting images, cells were segmented with Cellpose based on the pre-trained 'TN2' model[31]. For each segmented object, the position and mean intensity per (unaltered) channel were collected, and morphological descriptors ('eccentricity', 'perimeter', 'convex_area', 'area', 'axis_major_length', 'axis_minor_length') were extracted with scikit-image[70]. Cell profiles were then filtered with BioProfiling.jl[71], defining a set of filters (area < 1500 pixels, integrated nuclear intensity > 2000, average gold channel signal < 1000), and visually confirming that cell profiles were retained while artefacts were discarded.

**Intensity scaling**

We standardized marker abundance data from the cell feature table to make their distributions more comparable and more amenable to downstream analyses. To set channel intensities on comparable scales, we scaled intensity values in each channel by their 99.9th percentile. The resulting values were multiplied by 10, transformed with *arcsinh* and scaled by *arcsinh(10)* to make distribution shifts between background intensities and marker-positive cells more visible, while keeping 99.9% of the data between zero and one.

**Phenotyping**

We adopted a consensus approach, combining several methods to annotate cell types. First, we selected key lineage markers ("SMA", "CD31", "CD163", "CD68", "CD8", "CD45", "PanCK", "MPO", "CD7") and labelled each cell based on which one of these markers was the most abundant. To refine immune classification, cells with a dominant "CD45" abundance were relabelled based on immune markers ("CD3e", "CD8", "CD7", "CD14", "MPO", "CD20", "CD68", "CD163", "HLADR"), and cells dominated by "CD3e" were further relabelled based on lymphocytic markers ("CD4", "CD8", "CD7"). Then, we used PyFlowSOM (https://github.com/angelolab/pyFlowSOM), a fast Python wrapper of the FlowSOM method[72], to divide cells into 100 clusters, which were annotated based on the major marker labels we previously derived. Including categories for cells with ambiguous profiles, this led to 17 classes (*Major marker*, Supp. Fig. S2b).

Independently, a coarse annotation was obtained with Scyan[73]. Based on assumptions about five lineage markers ("CD14", "CD31", "CD45", "SMA", "PanCK"), cells were classified as lymphoid, myeloid, endothelial, fibroblast, epithelial (cancer) or "other", resulting in six classes (*Scyan*, Supp. Fig. S2b).

Finally, we looked at agreements between both sets of labels. If the annotations were coherent, the most precise description was kept. By visually inspecting the images, we also identified some systematic biases in annotation that could be corrected. For instance, cells initially labelled as NK or endothelial cells by our FlowSOM approach but as cancer cells by Scyan were indeed cancer cells. All the cases that could be systematically explained are summarized in Supp. Fig. S2b. Otherwise, cell annotations were considered unclear. Certain cell types were harder to profile than others due to their propensity to overlap with other cells (leading to signal contamination in the cell-level profiles) or to the markers included in our panel to identify them. For instance, NK cells could only be characterized by the presence of CD7 and CD45 and absence of T cell markers. Thus, we designed our phenotyping approach to focus on cells with a clear cell type, while $CD45^+$ cells that could otherwise not be identified were grouped as "other immune cells", and cells with low-confidence annotation were classified as "unclear" and not included in downstream analyses that required cell-type stratification.

**Cell type abundance comparison**

To compare the relative abundance of immune cell types per sample, and disentangle the compositional effects, we transformed the data as follows. First, zero frequencies were incremented by a small constant (1/10th of the smallest non-zero value), then proportions per sample were converted to centered log-ratios (CLR)[74]. For each observation (sample), this is computed as the natural logarithm of the original compositional values divided by their geometric mean.

The distribution of these values per clinical group was compared using a Mann-Whitney U test when comparing two groups (e.g. healthy samples against tumors) or a Kendall tau test when comparing multiple ranked groups (e.g. ordered pT stages). The statistics and associated two-sided p-values were computed using their scipy implementation[75], then corrected for multiple testing across immune cell types by computing a false-discovery rate with the Benjamini-Hochberg procedure.

**XGBoost modelling**

We evaluated how well measured features of CRC samples predict cancer stage, used here as a proxy for disease progression. Features were either cell-level (e.g. marker abundance) or sample-level (e.g. cell type abundance). To make inputs comparable, cell-level features were aggregated using their mean and standard deviation per sample. We modelled four classes (pT1 to pT4) using gradient-boosted trees implemented in XGBoost[76]. As pT is estimated by pathologists based on whole sections and this value is estimated from our measurements in $0.16mm^2$ of a single section, we do not expect high task performance, but any significant association would show the relevance of the input features in disease progression.

To select model hyperparameters and evaluate performance, we used a stratified nested cross-validation design. We set aside 20% of patients as an outer validation set for final evaluation. The remaining patients were split into four folds, each serving as a test set in the inner loop for hyperparameter tuning. Data from the same patient were confined to a single split, and every fold contained all stage categories. Performance was measured by the macro-F1 score to weight stages equally.

As a baseline, we generated 1000 random predictors per test by resampling labels from the training set (with replacement). Because model selection is based on inner folds, the chosen model is expected to outperform most random baselines on those folds by design. On the outer fold, evidence of informativeness comes from exceeding the average random baseline rather than the best random draws. Weak models may appear strong in one fold, but by regression to the mean will not sustain performance across folds. Thus, a model that outperformed baseline on the inner folds and also performs above the median random baseline on the outer fold reflects actual predictive power.

To interpret stage-specific predictions, we used SHAP, which defines an additive feature attribution model based on expectations across all possible feature coalitions. With the shap package[77], we computed SHAP values for each feature and class on the correctly predicted samples from the inner folds. Restricting interpretation to correct predictions ensures that the attributions reflect how the model arrived at valid decisions. We then aggregated SHAP values per feature and stage to identify the top contributors, providing a cohort-level view of which cellular or molecular features most strongly supported accurate stage assignments.

**Spatial modelling of cell types**

We used the MISTy framework[39] to model the spatial distribution of cell types. We defined two views: a *juxtaview*, giving equal importance to all neighbors up to 40 pixels (or 15.6 µm), and a *paraview*, weighing the influence of neighbors between 40 and 120 pixels (or 46.9 µm) following a Gaussian distribution. In each sample, Random Forest models[78] were used to predict the identity of each cell based on both views. The model performance was described separately for healthy and tumor samples. The information provided by the views is quantified as the $R^2$ percentage in predicting a cell type in a given sample. Additionally, the correlation coefficient is used to quantify whether a cell's identity is predicted from the presence or absence of a given surrounding cell type. For these analyses, the default MISTy parameters were conserved, only modelling cell types with positive $R^2$ (trim = 0) and censoring in group-level visualization the interactions with importance below 1 (cutoff = 1). Finally, the importance score of individual interaction patterns in the juxtaview or paraview, defined as the amount of variance reduction in the target expression centered and scaled, was used as the input features for tumor-stage predictive analyses. To denoise these features, any sample importance value below 0.9 was set to 0. The top 100 features with the most non-zero elements were used as candidate predictive features.

**Spatial modelling of metabolic markers**

To further look at the spatial organization of metabolic processes in the TME, we also modelled dependencies between metabolic markers in surrounding cells. We used the Kasumi approach[41], which expands the MISTy framework by modelling dependencies in subregions of each sample (sliding windows) to find patterns across scales. The view composition was kept as for the spatial cell type analysis, and the window size was set to 256 pixels (¼ of an image length, or 100 µm), with no overlap between windows. Windows with less than 20 cells were skipped. We set trim = 1 and cutoff = 0.9 to keep most of the spatial interactions detected in our analysis. However, to focus on the metabolic markers modelled the best, we restricted the intraview network to targets with $R^2 > 0.5$ (trim = 50). First, we summarized the results across all windows, then defined window clusters (cuts = 0.35 and res = 0.4), allowing us to identify interaction patterns that were restricted to subregions of the TME, while still being observed across multiple patients. The same parameters were used when describing cluster-specific patterns.

**Multicellular factor analysis**

We aimed to identify key axes of variation in the tumor microenvironment that would span the entire range of changes observable across cell types based on a spatial proteomics readout. For that, we built on the 'MOFAcell' approach that we adapted to our data and use case[42]. First, we selected fields of view with more than 20 epithelial cells. We grouped cell types to ensure that each population would be present in most samples: "APC", "B_cell", "Neutrophil" and "Other_immune_cell" were labelled as *Other_immune_cell*. "CD163_Macrophage" and "CD68_Macrophage" populations were combined as *Macrophage*. "CD4_Tcell" and "T_reg_cell" were grouped as *CD4_lymphocyte*. "NK_cell" and "CD8_Tcell" were labelled as *Cytotoxic_lymphocyte*. With this relabelling, no cell type was absent in more than 6% of the samples considered.

We computed the proportion of each of these eight broad cell type groups per sample. We then aggregated additional features for each of these cell types: We included the mean abundance of metabolic (CA9, CD98, CytC, MCT1, ASCT2, LDH, GS, GLS, ATP5A, CS, PKM2, GLUT1, ARG1, CPT1A) and functional markers (Ki67, MSH2, MSH6, STING1 and either PD-1 for T cells or PD-L1 for the rest) per cell type per image. We included the importance scores of the spatial patterns predicted by the presence of each cell type according to our previous MISTy analysis. The spatial metabolic predictors from Kasumi were not included, as they did not pertain to specific cell types but were only defined per sample. Finally, we included the mean and standard deviation of the morphological descriptors per cell type per sample.

All these measurements constitute a three-dimensional data structure, where one axis represents the cell type, one axis the sample, and one axis the measured feature (Fig. 5a). The number of features per cell type varied, as they included a variable number of spatial features. Via matrix factorization, the information can be approximated by the multiplication of a score matrix (latent factors by samples) and eight loading matrices (features by latent factors) for each cell type. The different cell types can be considered different views (or modalities) that can be simultaneously captured with a factor analysis framework[42,79]. Several frameworks are available to perform such factorization using variational inference, differing based on the chosen priors. With a set number of 10 factors, the unweighted cell type average reconstruction Pearson correlation coefficient was higher for MuVI[43] with signed factors (0.63) than for MuVI with non-negative factorization (0.59) or for MOFA+ (0.58), hence this approach was selected for further analyses.

We identified that factors 2 and 6 were associated with the pT stage (FWER $< 0.05$, Kruskal-Wallis rank sum test). Thus, we considered that the features with the highest absolute loadings for these factors were defining a solid base to describe disease progression. For each cell type, we selected the five features with the highest absolute loadings in one of these factors (Supp. Table 1). We then adopted a cross-validation scheme to test the predictability of progression-related features across cell types. The sample distribution was kept as for the pT stage predictions we previously performed, with additional exclusion in each split of samples lacking features in the test set (i.e. samples that did not contain sufficient cells of the corresponding cell type). We used XGBoost regression models with a squared error objective function. For each pair of cell types, the number of estimators and maximal tree depth were selected on the inner cross-validation folds, then the performance (reconstruction $R^2$) was assessed on the holdout validation fold. For computational efficiency, we skipped the pairs which had a median $R^2$ score $< 0.2$ on the model selection folds. Finally, networks representing the predictability of progression-associated features across cell types were compiled by defining $R^2$-weighted edges for all cell type pairs with a validation score $> 0.2$, stratifying the holdout set between early (pT1, pT2) or late (pT3, pT4) validation samples.

A similar analysis was conducted by extracting the top five features with the highest absolute loadings in factors 2 and 6, grouped by feature type (morphological, functional, metabolic, spatial, abundance) instead of cell type. However, this did not result in any pair of feature types with prediction performance $R^2 > 0.2$.

**MuVIcell Python package**

To facilitate the re-use of the workflow we developed to other multicellular profiling datasets, we compiled all the relevant preprocessing steps, helper functions and visualizations into the *muvicell* Python package. The code is openly available and shared with a GPL-3.0 license (https://doi.org/10.5281/zenodo.17186802). For reproducibility, we define package dependencies to allow convenient installation using uv or conda.

The package guides users to start from anndata measurements and aggregate them per cell type, or directly start from muon cell-type stratified objects, then normalize the data before running MuVI. The resulting factors are interrogated using a set of helper and visualization functions to identify what features drive each factor and quantify the involvement of the different cell types. These factors can also be tested for statistical association with clinical variables such as pathological classes or mutational status, if available.

**Transcriptomics validation**

Cell type abundance per disease stage was directly obtained from Joanito et al[51].

For the reanalysis of the data from Pelka, Hofree, Chen et al[9], we started from the provided Loupe file of epithelial cells. With Loupe Browser 8 (10x Genomics), we first selected cells with a good coverage (*RPLP0* and *PPIA* counts > 5) and some degree of proliferation (*MKI67* counts > 2). We defined cells with at least two counts of both *LY6E* and *TGFBI* to be malignant, and cells with no count of either to be healthy. We then extracted the abundance of metabolic genes of interest for these populations. *ARG1* was not included because of low counts. We used the built-in tool to perform a differential analysis between the gene expression levels in both populations. Using the ranked differential expression list, we looked for enrichments with DecoupleR[80] in pathway transcriptional footprints (PROGENy[57]) and regulatory motifs (CollecTRI[81]).

To validate our progression-related spatial findings, we explored the data generated by Heiser et al[8]. Starting from the pre-processed spot-level data, we followed the notebook provided to compile lineage signatures. We further added signatures for signaling footprints with PROGENy[57] and metabolic pathways (OXPHOS, FAO, Glycolysis) with MSigDB[58], based on read-count-invariant Univariate Linear Model enrichments from DecoupleR[80].

To compare relative preference of energy metabolism pathways, we transformed the data as follows: We computed median metabolic signatures (OXPHOS, FAO, Glycolysis) per sample, scaled each signature between 0 and 1, then scaled the sum of all three signatures per patient to sum up to 1. With this transformation, a sample with a median activity of each pathway would be scored as (⅓, ⅓, ⅓) while a sample with increased glycolytic activity and lowered activity of OXPHOS and FAO would get a score tending towards (1,0,0). As a null hypothesis (uniform probability to observe any metabolic preference), we considered that observations would be sampled from a Dirichlet distribution with parameter (1,1,1). To test FAO enrichment in early-grade tumors, we compared the mean distance to FAO vertex compared to shuffled labels.

Finally, we aimed to test the predictability of tumor grade (G1, G2 or G3) based on spot-level lineage signatures. As input, we used the median levels across spots per sample of epithelial-related or TME-related (stromal and immune) signatures, as classified in the original publication. A total of 47 samples with known tumor grade were available, stemming from 29 distinct patients. Due to the limited sample size, we adopted a simple leave-one-patient-out cross-validation scheme, ensuring all grades were present in both training and testing sets, and did not perform extensive hyperparameter optimization. We report results for XGBoost modelling with 20 estimators and a maximum depth of 5, and we noted that the results were robust to changes in these parameters. Macro-F1 scores represented the overall performance of the model in predicting tumor grade across all samples. A subset of 15 patients had multiple samples and was also used to compute prediction accuracy, which quantified the robustness of the models across iterations.

**Use of large language model assistants**

During the preparation of this work the authors used Perplexity (Perplexity AI), NotebookLM (Google) and ChatGPT (OpenAI) in order to perform extended literature search and assist with copy editing and text clarity. The authors reviewed and edited the content as needed and take full responsibility for the content of the published article.

**Figures and illustrations**

Illustrations in Fig. 1 Fig. 2h, Fig. 3a, Fig. 6j and Supp. Fig. S3a include elements created with BioRender (https://app.biorender.com/). Complete figures were assembled and edited with Affinity Designer 2.

**Code availability**

All the scripts used in this analysis can be found on GitHub, along with instructions on how to reproduce the results reported in this manuscript:
https://doi.org/10.5281/zenodo.17008988

## 5. Acknowledgements

This publication was supported through state funds approved by the State Parliament of Baden-Württemberg for the Innovation Campus Health + Life Science alliance Heidelberg Mannheim. This project has received funding from the European Research Council (ERC) under the European Union's Horizon 2020 research and innovation program (grant agreement 101116823) and from the Helmholtz association (VH-NG-1605), both to F.J.H.
We thank Ricardo O. Ramirez Flores, Christina Schmidt, Charlotte Boys, Philipp S. L. Schäfer and David R. Glass for their valuable feedback and suggestions.

## 6. Conflict of interests

JSR reports in the last 3 years funding from GSK and Pfizer & fees/honoraria from Travere Therapeutics, Stadapharm, Astex, Owkin, Pfizer, Grunenthal, Tempus and Moderna.

## 7. Authors' contributions

Conceptualization: L.V., T.G., G.S., F.J.H.
Data curation: L.V., T.G.
Formal analysis: L.V.
Funding acquisition: J.S.R.,F.J.H.
Investigation: L.V., T.G., S.T., M.C.
Methodology: L.V., J.T.
Project administration: L.V., T.G, F.J.H.
Resources: L.B., R.S., G.S.
Software: L.V., T.G., Y.W.
Supervision: J.S.R., F.J.H.
Validation: L.V.
Visualization: L.V.
Writing – original draft: L.V., S.T., M.C., Y.W., F.J.H.
Writing – review & editing: all authors.

## Supplementary Materials

### Supplementary Tables

All supplementary tables are available on Zenodo (https://doi.org/10.5281/zenodo.17280877)

### Supplementary Figures

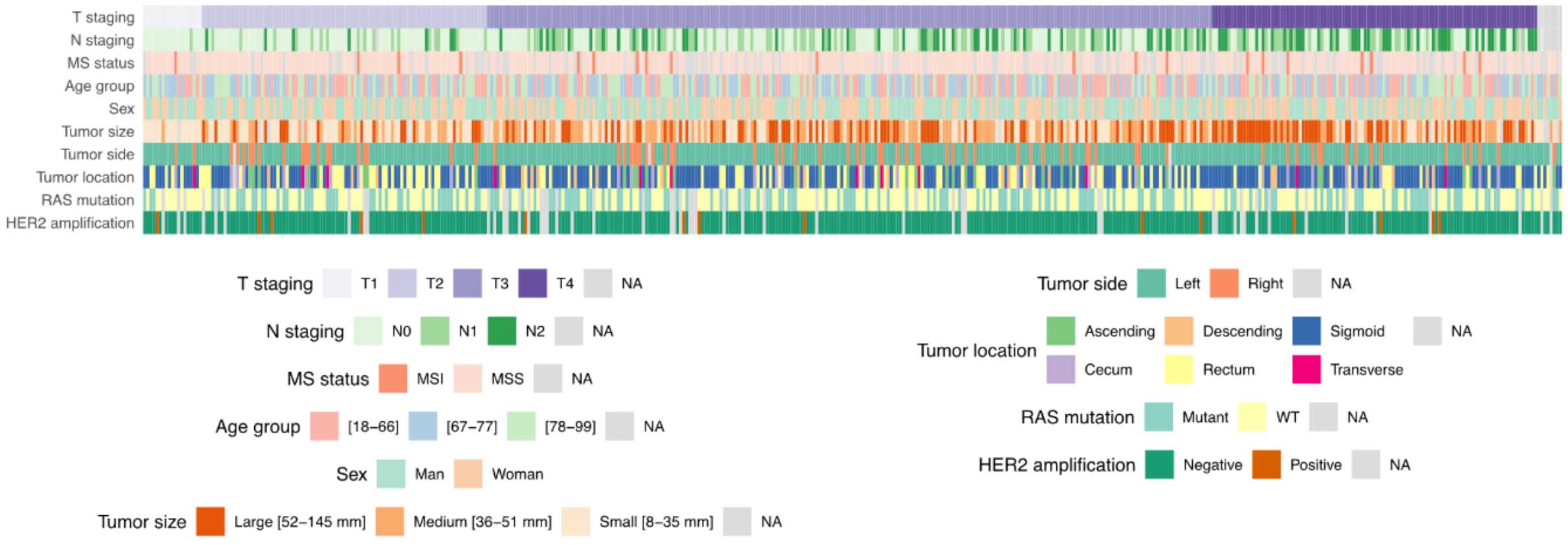

**Supplementary Fig. S1: CRC cohort profiled.**

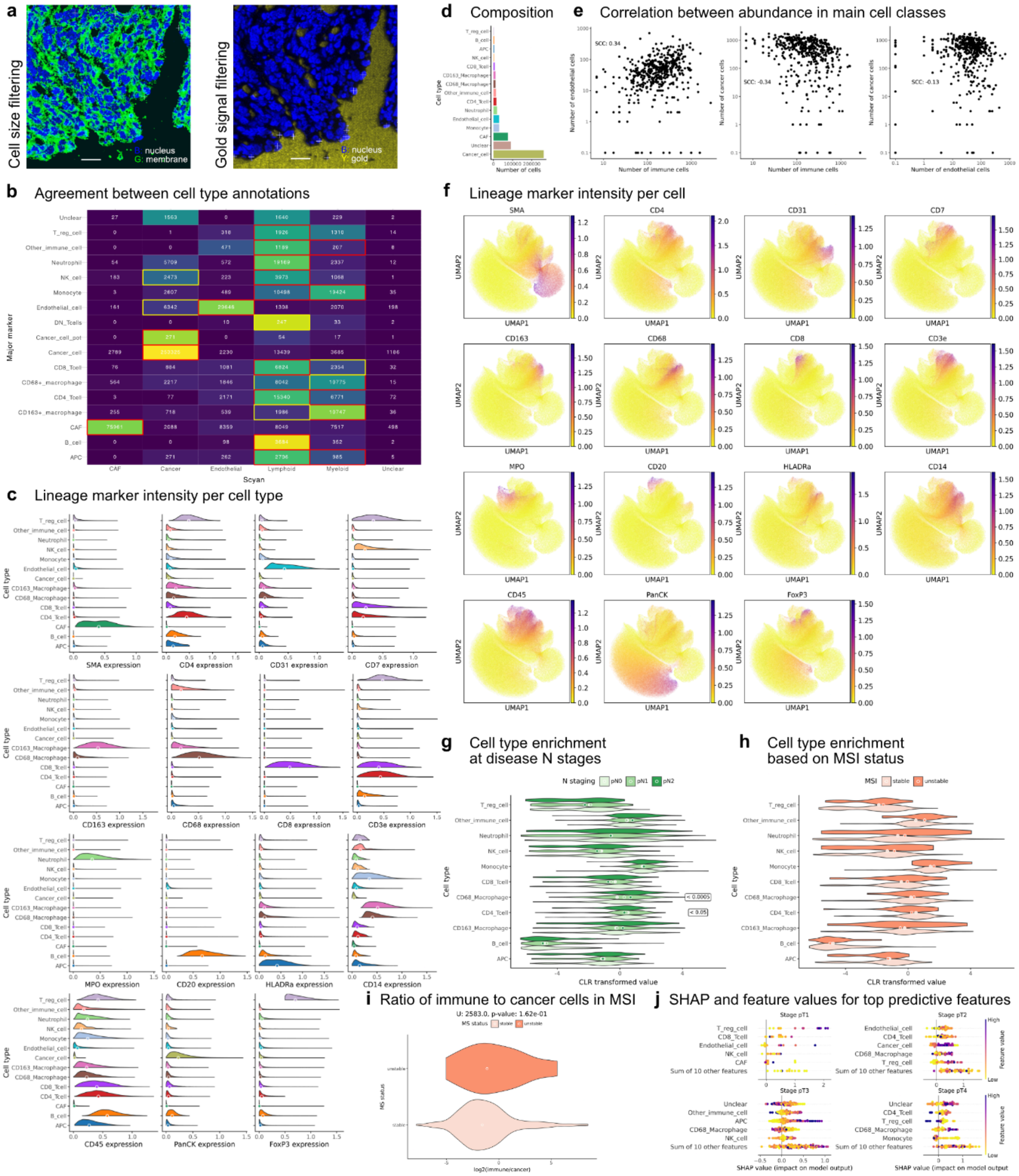
a
Cell size filtering
B: nucleus
G: membrane
Gold signal filtering
B: nucleus
Y: gold
d
Composition
e
Correlation between abundance in main cell classes
b
Agreement between cell type annotations
Major marker
Scyan
f
Lineage marker intensity per cell
c
Lineage marker intensity per cell type
g
Cell type enrichment
at disease N stages
h
Cell type enrichment
based on MSI status
i
Ratio of immune to cancer cells in MSI
U: 2583.0, p-value: 1.62e-01
j
SHAP and feature values for top predictive features

**Supplementary Fig. S2: Properties of the cell lineages profiled.**

**a** Example of objects filtered out based on their surface area or their signal in the gold channel (slide background) in a cropped field of view. Each discarded object is identified by a white cross at its center. **b** Agreement of cell type annotations obtained based on either the highest lineage marker intensity ("Major marker", rows) or on a probabilistic model of multiple lineage marker intensities ("Scyan", columns). After curation, cases highlighted in red were mapped to their "Major marker" label, cases highlighted were mapped to a different, corrected label, and all other combinations were labelled as "Unclear" (see Methods for details). **c** Distribution of the lineage markers for each cell type. **d** Number of resulting cell annotations across images. **e** Relation between number of immune, endothelial and cancer cells in all samples. SCC: Spearman Correlation Coefficient. **f** UMAP embedding showing the distribution of lineage marker intensities per cell in a shared lineage space. Corresponding cell types can be visualized in Fig. 2d. **g-h** Comparison of cell types present at different node infiltration stages (g) or in tumors with different microsatellite stability statuses (h). Values are transformed to account for the compositional property of cell type abundances and make them independently comparable, and FDR values $< 0.05$ are reported for Kendall's $\tau$ and Mann-Whitney U tests, respectively (see Methods for details). CLR: centered log-ratio. **i** Ratio between absolute number of all immune cells and cancer cells per patient in MSI and MSS samples. Differences were tested using the Mann-Whitney U test. **j** Association between original feature values and SHAP values per stage for top predictive compositional features. Central dots represent median values in (c), (g), (h), and (i).

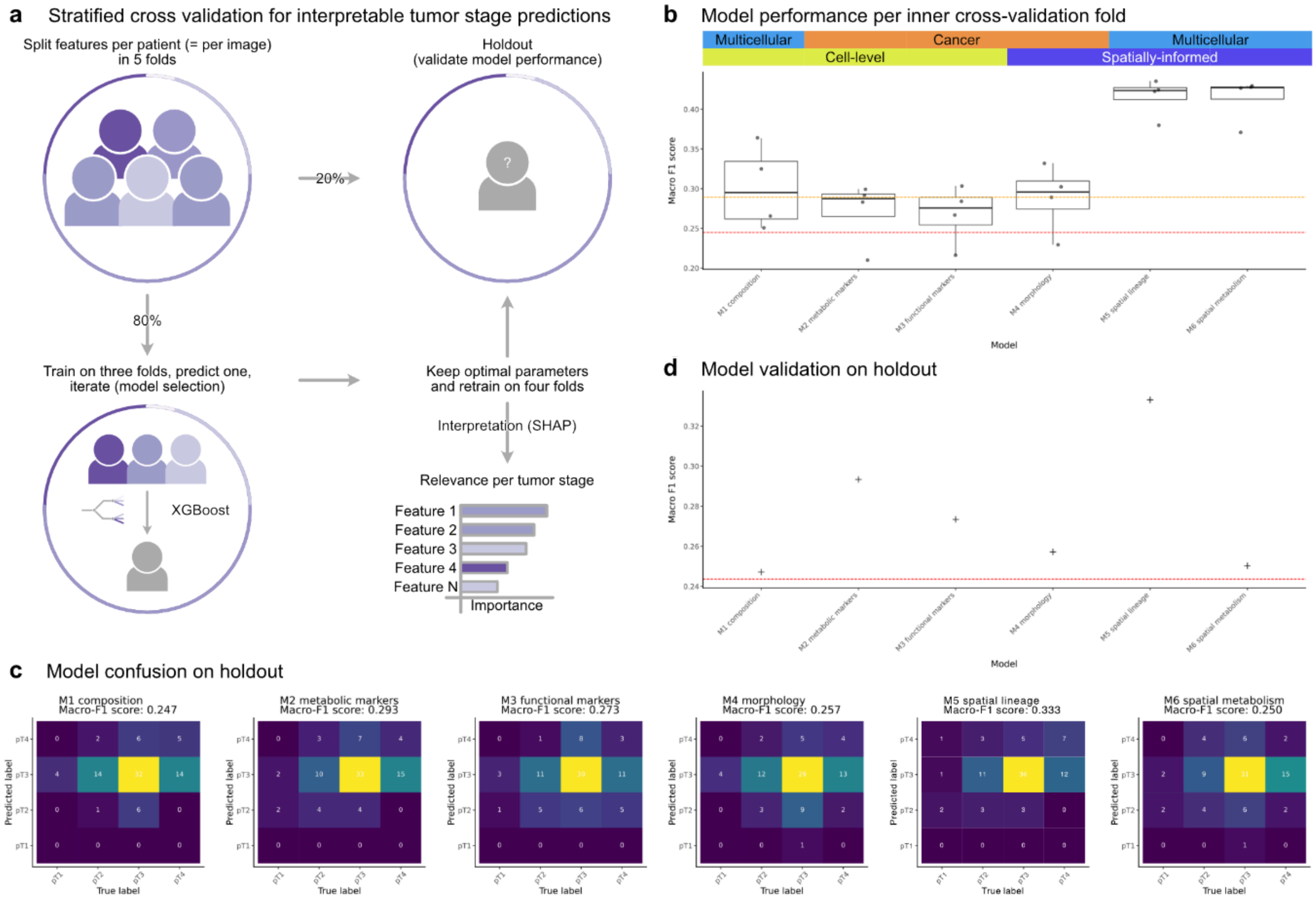

**Supplementary Fig. S3: Predictive machine learning approach highlighting features associated with tumor stage.**

**a** Overview of predictive modelling performed to map molecular and spatial features to tumor stages across patients. XGBoost: extreme gradient boosting. SHAP: Shapley Additive Explanations. **b** Macro-F1 score per fold of the inner cross-validation loop used to select robust models. Orange dotted line represents the 95th-percentile of the median score across folds for the baseline model (see Methods), and red dotted line represents the median of these values. **c** Confusion matrices on held-out validation data for each model considered. **d** Summary of the performance of each model on held-out validation data. Red dotted line displays the median F1-score for the validation samples obtained using the baseline approach.

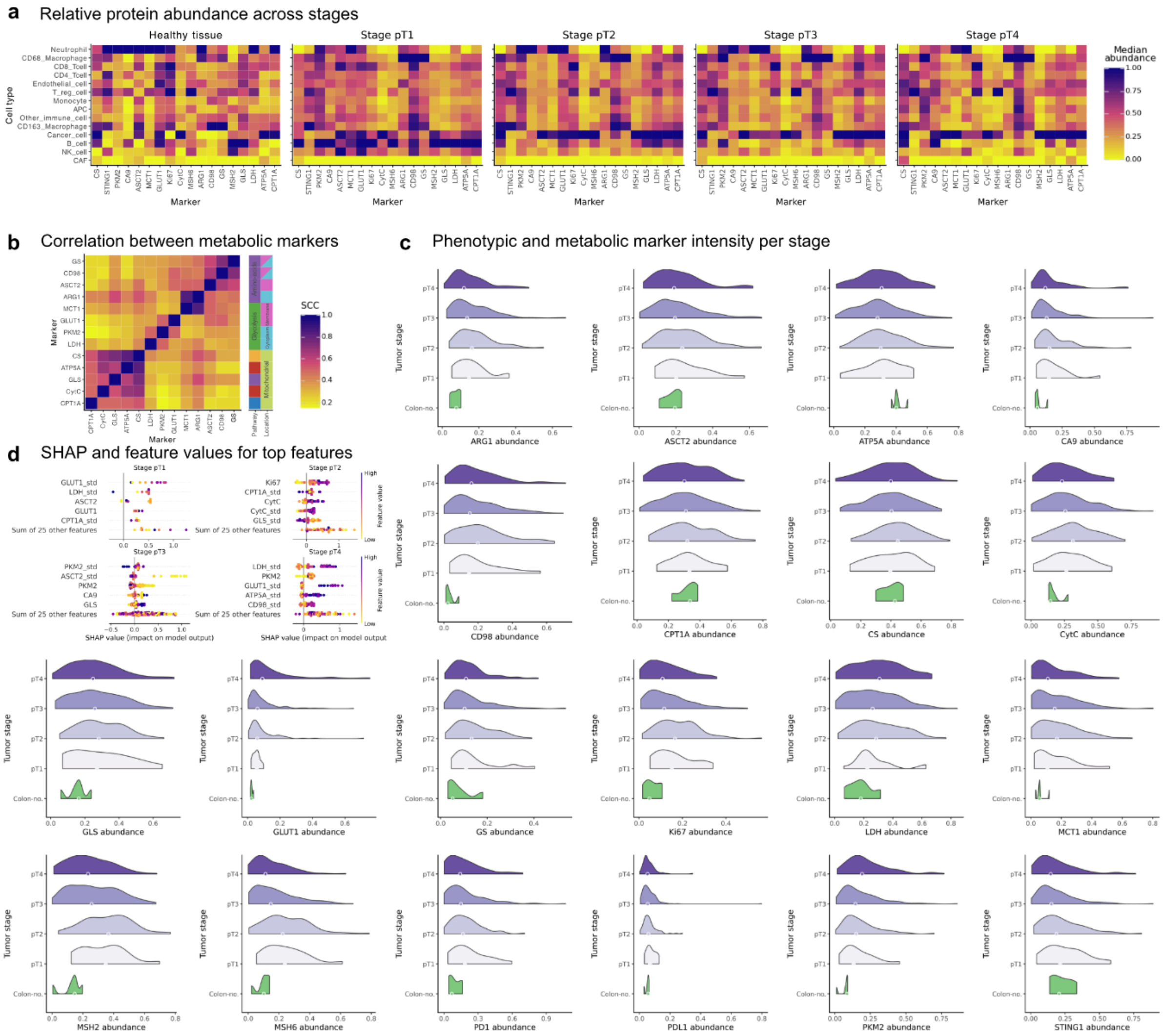

**Supplementary Fig. S4: Metabolic profiles and clustering in epithelial cells.**
**a** Median metabolic and functional marker abundance per cell type, stratified by disease status and cancer primary tumor staging. **b** Correlation between metabolic markers in cancer cells. Corresponding metabolic pathways are colored as in ***Fig. 3***a and subcellular location was fetched from the Human Cell Atlas (version 24.0)[82]. SCC: Spearman Correlation Coefficient. **c** Distribution of the metabolic and functional markers for each tumor stage. Central dots represent median values. Central dots represent median values. **d** Association between original feature values and SHAP values per stage for top predictive metabolic features.

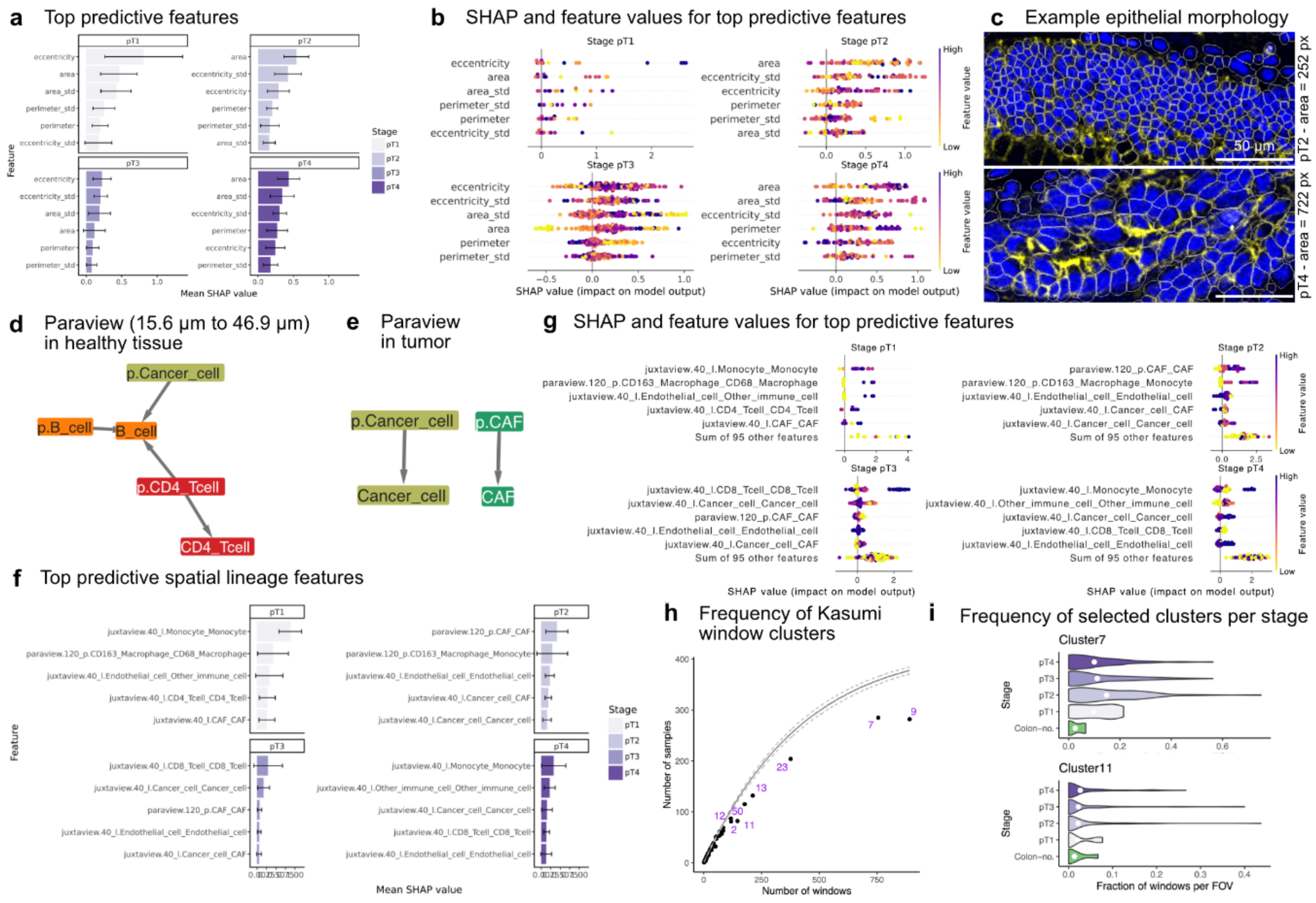

**Supplementary Fig. S5: Detailed spatial and morphological characterization of samples.** **a** Mean importance (SHAP values) of top predictive morphological features per stage. **b** Association between original feature values and SHAP values per stage for top predictive features. **c** Example images displaying tumor cells with distinct morphology. Yellow = cytokeratin, blue = CLAHE-corrected nuclear channel, white outline = segmented cells. The pT stage and median cancer cell area corresponding to each image are reported.
**d-e** Network of cell type associations in indirect neighborhoods in healthy (d) and tumor tissue (e). **f** Mean importance (SHAP values) of top predictive spatial lineage features per stage. **g** Association between original feature values and SHAP values per stage for top predictive features. **h** Frequency of metabolic regulation clusters per sample and per window. Solid gray and dotted lines represent the median and 80% confidence interval of the distribution of the expected relation between number of windows and samples in which a cluster would be present, if following a multivariate hypergeometric distribution. All frequent clusters are not randomly spread across samples but more concentrated than expected given their abundance. **i** Frequency of clusters 7 and 11 per image in healthy samples and at different tumor stages. Central dots represent mean values.

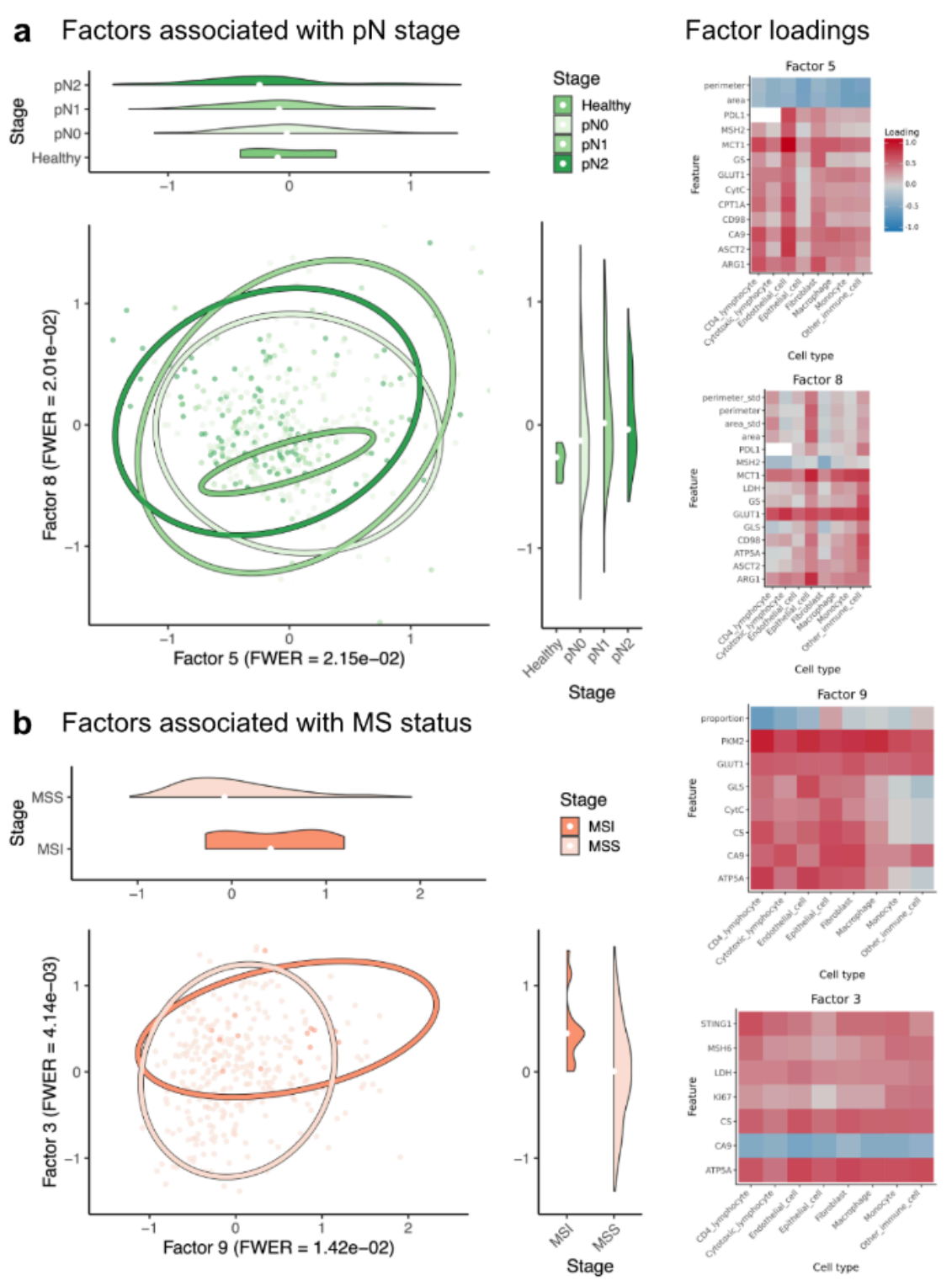

**Supplementary Fig. S6: Clinical variables associated with multicellular factors.**
**a-b** Multicellular factors associated with pN stage (a) and MS (b). Left: Confidence ellipses show the region within 2 standard deviations from the mean per clinical variable level. Ranges are selected to best show the ellipses, despite outlier samples being truncated. Central dots represent median values. Right: Loadings of top features per factor.
MS: microsatellite stability.

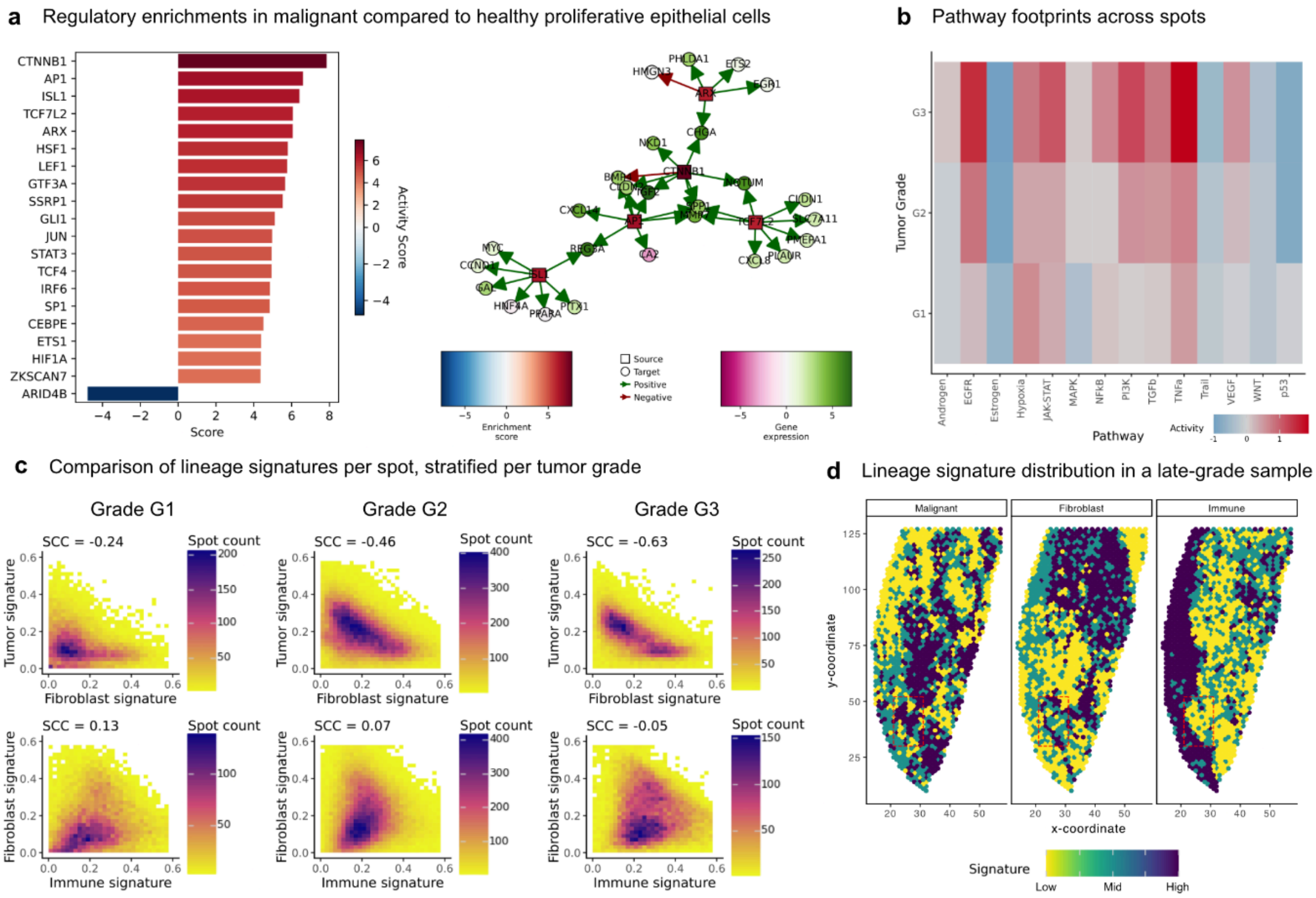

**Supplementary Fig. S7: Mechanistic insights from transcriptomics CRC profiles.**

**a** Transcription factor target enrichments based on the CollecTRI curated set[81] (left) along with the corresponding regulation network (right). Up to 10 regulons were visualized for the four TFs with the highest inferred activity. **b** Enrichment in PROGENy signaling footprints[57] per tumor grade. **c** Density maps of the relation between fibroblast and tumor or immune signatures per spot stratified by tumor grade, along with the corresponding Spearman Correlation Coefficient (SCC). d Example of binned normalized lineage signatures in a grade 3 sample. Area delineated by red dotted lines highlight a region where a fibroblast layer was separating epithelial from immune cells. Data from Pelka, Hofree, Chen et al[9] in (a). Data from Heiser et al[8] in (b-d).